\documentclass[aps,prb,twocolumn,floatfix,amsmath,amssymb,showpacs,
               superscriptaddress,10pt]{revtex4-1}
\usepackage[final]{graphicx}
\usepackage{placeins} 
\usepackage{float}
\usepackage{color}
\usepackage{dcolumn}
\usepackage{caption,subcaption}
\usepackage{xcolor}
\usepackage{url}
\graphicspath{{../Article/Submission/figures/}}
\usepackage{bm}
\usepackage{epsfig,psfrag,amsmath,amssymb}
\input{epsf}
\usepackage{hyperref}
\usepackage[percent]{overpic}

\hypersetup{
    colorlinks=true,
    linkcolor=blue,
    citecolor=blue,
    filecolor=magenta,      
    urlcolor=blue,
}
\usepackage{array}
\usepackage{tabu}
\newcolumntype{C}[1]{>{\centering\arraybackslash$}p{#1}<{$}}

\setcitestyle{square}
\newcommand{\pdag}{{\phantom{\dagger}}}
\bibpunct{[}{]}{,}{n}{}{}
 
\begin{document}
\title{Flat Bands and Ferrimagnetic Order in \\
Electronically Correlated Dice-Lattice Ribbons}

\author{Rahul Soni}
\affiliation{Department of Physics and Astronomy, The University of 
Tennessee, Knoxville, Tennessee 37996, USA}
\affiliation{Materials Science and Technology Division, Oak Ridge National 
Laboratory, Oak Ridge, Tennessee 37831, USA}
\author{Nitin Kaushal}
\affiliation{Department of Physics and Astronomy, The University of 
Tennessee, Knoxville, Tennessee 37996, USA}
\affiliation{Materials Science and Technology Division, Oak Ridge National 
Laboratory, Oak Ridge, Tennessee 37831, USA}
\author{Satoshi Okamoto}
\affiliation{Materials Science and Technology Division, Oak Ridge National 
Laboratory, Oak Ridge, Tennessee 37831, USA}
\author{Elbio Dagotto}
\affiliation{Department of Physics and Astronomy, The University of 
Tennessee, Knoxville, Tennessee 37996, USA}
\affiliation{Materials Science and Technology Division, Oak Ridge National 
Laboratory, Oak Ridge, Tennessee 37831, USA}
\date{\today}

\begin{abstract}
We study ribbons of the dice two-dimensional lattice (that we call ``dice ladders'') known to have
nontrivial topological properties, such as Chern numbers 2 [Wang and Y. Ran, Phys. Rev. B {\bf 84}, 241103 (2011)].
Our main results are two folded: (1) Analyzing the tight-binding model in the presence
of Rashba spin-orbit coupling and an external magnetic field, we observed that dice ladders qualitatively display
properties similar to their two-dimensional counterpart all the way to the limit of only two legs 
in the short direction. This includes flat bands near the Fermi level, edge currents and edge charge
localization near zero energy when open boundary conditions are used, two chiral edge modes, 
and a nonzero Hall conductance. (2) We studied the effect of Hubbard correlation $U$ in the two-leg dice ladder
using Lanczos and density matrix renormalization group techniques.
We show that increasing $U$ the flat bands split without the need of introducing external fields. Moreover,
robust ferrimagnetic order develops. Overall, our work establishes dice ladders as a promising playground 
to study the combined effect of topology and correlation effects, one
of the frontiers in Quantum Materials.
\end{abstract}
\maketitle

\section{Introduction and Dice Lattice}

In his seminal paper, Haldane proposed a tight-binding model on a honeycomb lattice, including a staggered flux pattern, that displays the integer quantum Hall effect~\cite{haldane88}. Time reversal symmetry is broken by the complex-valued second-neighbor hoppings and no external magnetic fields are needed. Generalizations led to the concept of topological insulators~\cite{top-ins1,top-ins2,top-ins3}. Later, it was discovered that a similar approach works even for the fractional quantum Hall effect, also without employing Landau levels and large magnetic fields~\cite{FQHE1,FQHE2,FQHE3,FQHE4,FQHE5}. The essence is to search for models that display quasi two-dimensional flat bands with a nonzero Chern number.
The topological band structures created by this procedure are of considerable current interest. This type of models are referred to as lattice Chern insulators. The Chern number equals the number of chiral edge modes.

\begin{figure}[H] 
\centering
\includegraphics[width=6in, height=2in]{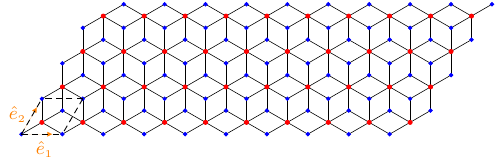}
\caption{Dice lattice geometry. Blue dots are sites with coordination number 3, while red dots have coordination number 6. The dashed black box describes the unit cell of the lattice. $\hat{e}_{1}$ and $\hat{e}_{2}$ (in orange) are the lattice unit vectors. Specifically, here we show a 10$\times $4 lattice with 10 unit cells along the $\hat{e}_{1}$ and 4 unit cells along the $\hat{e}_{2}$ directions (simply count 
the number of red dots).\label{fig1}}\label{Geometry of the lattice}
\end{figure}

The two-dimensional Haldane model involving two triangular lattices was generalized to three by Wang and Ran~\cite{wang11}. By this procedure fine tuning of parameters is avoided, and the honeycomb lattice becomes the {\it dice} lattice~\cite{dice-previous1,dice-previous2,dice-previous3}, Fig.~\ref{Geometry of the lattice}, via the addition of an extra site at the center of each hexagon. This bipartite lattice has two types of sites: two thirds with coordination three and one third with coordination six. The unit cell contains three sites, thus leading to three bands. At the non-interacting tight-binding level and with a small Zeeman field -- with hoppings only between nearest-neighbor sites and including Rashba spin-orbit coupling $\lambda$ -- the model displays bands with Chern number $C=2$~\cite{cook14}, as opposed to the more standard $C=1$ of Landau levels in the integer quantum Hall effect.

The dice lattice could be realized in cold atoms or via a trilayer superlattice grown in the [111] direction~\cite{okamoto18}, such as in SrTiO$_3$/SrIrO$_3$/SrTiO$_3$. From the theory perspective, the $\alpha$-$T_3$ model, interpolating between the honeycomb and dice lattices, also received considerable attention~\cite{illes15,illes16,islam17,dey18}. In dice lattices, the effect of interactions was considered using weak coupling and mean-field techniques~\cite{dora14}. Its realization in optical lattices~\cite{moller12,andri15} and in LaAlO$_3$/SrTiO$_3$ (111) quantum wells~\cite{doe13} was also studied. Bulk oxides with the generic
formula A$_4$B$^{\prime}$B$_2$O$_{12}$, such as Ba$_4$CoRe$_2$O$_{12}$~\cite{zhou}, contain trilayers that
seen ``from above'' also resemble a dice lattice.

\begin{figure}[H] 
\begin{subfigure}[b]{0.3\textwidth}
\centering
\includegraphics[width=2.8in, height=1.8in]{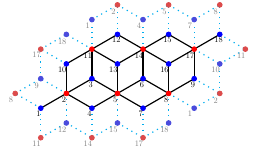}
\subcaption{ }
\end{subfigure}
\hfill
\begin{subfigure}[b]{0.3\textwidth}
\centering
\includegraphics[width=2.4in, height=1.8in]{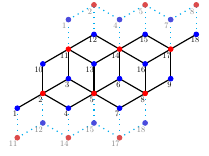}
\subcaption{ }
\end{subfigure}\quad
\begin{subfigure}[t]{0.3\textwidth}
\centering\raisebox{1.3cm}{
\includegraphics[width=2.4in, height=1.2in]{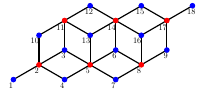}}
\vspace*{-1cm}
\subcaption{ }
\end{subfigure}
\caption{Boundary conditions used in our study, employing a 3$\times$2 unit-cells cluster as example. Each unit cell has 3 sites, thus 
there are 18 sites in this cluster. (a) corresponds to PBC$\times$PBC i.e. periodic boundary conditions in both directions. The specific neighbors of each site at the boundary
are labeled by site numbers. The solid (dashed) lines denote connections inside the cluster (across the boundary). (b) Similar as (a) but with PBC along the vertical
direction and OBC along the horizontal direction. (c) Similar to (a) but using open boundary conditions in both directions. Here ``open'' refers to a sharp edge, as opposed to connection to leads or reservoirs.\label{fig1-extra}}\label{boundary-conditions}
\end{figure}

In this publication, we study ribbons of the dice lattice. This is equivalent to a dimensional reduction from two to one of the original dice lattice into a quasi-one dimensional system, conceptually similar to studying graphene in two dimensions vs carbon nanotubes in one dimension: locally the atomic connections are the same, but globally both systems are different. Graphene nanoribbons were also studied in recent efforts~\cite{louie}.
We refer to the dice ribbons as ``dice ladders'' borrowing the language of correlated electronic systems: the dice lattice is two dimensional, while the dice ladder is one dimensional, with a finite length in the short direction. We use three primary boundary conditions: ({\it i}) cylindrical geometry, with periodic boundary conditions (PBC) in the short direction and open boundary conditions (OBC) in the long one, or viceversa; ({\it ii}) torus geometry, with PBC in both directions; ({\it iii}) open geometry, with OBC in both directions (i.e. sharp edges). 
The latter is important to visualize in real space the edge currents and edge charge localization. 
Twisted boundary conditions and associated average over many edge angles 
were not used, because such procedure would be too demanding in CPU time when including a Hubbard $U$. 
Because the dice lattice has a complicated geometry, and the unit 
cell has three sites, for the benefit of the reader the case of a 3$\times$2 unit-cells cluster with the three boundary conditions used in our work is illustrated in Fig.~\ref{boundary-conditions}. Note that this cluster has a total of 18 sites. Also, to avoid confusion with alternative conventions used in Mesoscopic physics, note that we follow the notation of correlated electrons where ``open boundary conditions'' means that the system ends abruptly at the edge. Open here is not used in the sense of connected to, for example, leads.

\onecolumngrid

\twocolumngrid

\onecolumngrid

\twocolumngrid

Our effort has unveiled two primary results: (1) Dice ladders, i.e. dice ribbons, keep their primary properties even in the smallest dice ladder we studied. (2) Exploiting this result, numerically we introduced correlation Hubbard $U$ effects and found that ferrimagnetic order develops with increasing $U$. More specifically:

(1) Using the three boundary conditions described above, we unveiled a rich eigenvalue spectrum that, surprisingly, continues displaying flat bands even into the smallest ladders we have studied. In principle flat bands originate in localized states and reducing one dimension should have minor consequences until the natural localization length scale is reached. What is interesting of our results is that the flat bands survive even for the smallest ladder we studied, the $N \times 2$ ladder. It was not possible to anticipate {\it a priori} that such a thin system would still have flat bands. In addition, all the geometries studied here display a nonzero Hall conductance in the
flat bands region, with the
particular value of the 4-leg ladder system at the Fermi level being very close to $C=2$ as in the planar dice lattice. 
Moreover, edge currents and edge charge localization appear in all the ladders near zero energy, 
even in the smallest $N \times 2$ cluster, when using both OBC$\times$OBC 
and PBC$\times$OBC. Thus, 
even the smallest dice ladder studied here displays properties that resemble the two-dimensional topologically nontrivial system~\cite{guo12,chen19}. Note that sharp topological numbers, such as integer 
Chern numbers, cannot be obtained using dice ladders because
of the finite number of momenta in the short direction when using cylindrical or fully periodic boundary conditions
that prevents a smooth definition of the momentum derivatives involved in the
integral that defines $C$. However, we will show that all properties of the 
quasi-one-dimensional dice ladders are qualitatively similar to those of the truly topological
two-dimensional systems, and in these regards dice ladders are useful ``toy models'' for dice planes. Their primary importance lies in the fact that correlation effects can be accurately introduced computationally in $N \times 2$ ladders but not in planar geometries.

(2) The influence of correlation effects on topological systems~\cite{rachel18} is an important open problem that can be studied in dice ladders. In particular, here in the $N \times 2$ ladder we incorporated Hubbard $U$ repulsion via powerful unbiased numerical techniques beyond mean-field approximations, such as the Lanczos~\cite{lanczos} and density matrix renormalization group (DMRG) methods~\cite{DMRGreview}. Below we show that as $U$ increases, ferrimagnetic order develops, in agreement with spin-1/2 Heisenberg model calculations~\cite{heisenberg}. Moreover, the original $U=0$ flat bands split due to the magnetic order that emerges. Then, for this split to occur external magnetic fields are not needed. Our results prove that dice ladders can be studied numerically with accuracy even with nonzero $U$ and opens several avenues of future investigation. How are the edge states affected by $U$? What is the complete phase diagram in the plane $U$-$\lambda$? Are there new phases? 

This path of future research resembles the successful strategy followed in cuprates and pnictides/selenides superconductors where also the two dimensional original problem was transformed to the geometry of
$N$-leg ladders to allow for computational accuracy. 
In particular, recent experimental and theoretical work showed that real two-leg ladder iron-based  materials, such as 
BaFe$_2$S$_3$~\cite{NatMatSC,PatelBaFe2S3}, turn superconducting with high pressure and 
display nontrivial magnetic properties~\cite{FeLadder7,NeutronOSMP,herbrych}. In Cu-oxide ladders, similar
success was theoretically and experimentally achieved~\cite{ladder1,ladder2,uehara96}.
Exploring the physics of interacting $N$-leg dice ladders
can define an equally fertile area of research. 

%

The organization of the manuscript is as follows. In Sec.~II, the model 
and methodology used will be introduced, as well as the
main observables calculated. Section III contains the main numerical results for the tight-binding 
model reduction from planes to ladders. We study the density-of-states (DOS) and flat bands, 
starting with the $N \times N$ cluster, followed by the
$N \times 4$ ladder, and then the $N \times 3$ and $N \times 2$ ladders. Section IV addresses the edge currents, edge charge localization, and Hall conductance, still within the non-interacting tight-binding model. Section V presents our results incorporating Hubbard $U$ correlations for the $N \times 2$ case, focusing on the ferrimagnetic order that develops and on the splitting of the flat bands even without magnetic fields.
Conclusions are in Sec.~VI. 

\section{Models and Methods}

\subsection{Hamiltonian}
In Sec. III and IV of this manuscript, we will consider the same non-interacting Hamiltonian studied in Ref.~\cite{wang11}. In this previous work, a two-dimensional geometry was used while here we use a ribbon geometry. 
This tight-binding Hamiltonian has three components: $H=H_{K} + H_{SOC} + H_{B}$, where $H_{K}$ is 
the tight-binding kinetic energy, 
$H_{SOC}$ is the Rashba spin-orbit coupling, and $H_{B}$ the interaction of electrons with an external magnetic field.  
The kinetic term  $H_K$ is defined as
\begin{equation}
H_{K} = -t\sum_{\substack{\mathbf{r},\alpha,\mathbf{r}',\beta\\ \sigma}}\left(c^{\dagger}_{\mathbf{r},\alpha,\sigma}c_{\mathbf{r}',\beta,\sigma}+h.c.\right) -\epsilon\sum_{\mathbf{r}} n_{\mathbf{r},2},
\end{equation}\label{Eqn: Kinetic term}

\noindent
where $\mathbf{r}, \mathbf{r}'$ are the unit cell indexes [$\mathbf{r}=(r_{1},r_{2})$ is a vector with
components in a coordinate system defined using the
unit vectors in Fig.~\ref{fig1}], $\alpha$ and $\beta$ are the site indexes within the unit cell $\mathbf{r}$ and $\mathbf{r}'$, respectively (with $\alpha,\beta=1,2,3$), and $\sigma=\uparrow , \downarrow$ is the $z$-axis spin projection of the electron 
at site $\alpha$ within the unit cell $\mathbf{r}$. The on-site energy is $\epsilon$ and $n_{\mathbf{r},\alpha}=\sum_{\sigma}c^{\dagger}_{\mathbf{r},\alpha,\sigma}c_{\mathbf{r},\alpha,\sigma}$ is the local density of electrons at site $\alpha$ in the unit cell $\mathbf{r}$. $\epsilon$ affects 
only the ``red'' sites of Fig.~\ref{fig1} (those with coordination 6).
This kinetic term depicts the inter- or intra-unit cell electronic tunneling between nearest-neighbor lattice sites $\mathbf{r},\alpha$ and $\mathbf{r}',\beta$ with hopping amplitude $t$. In this article, $t$ is the energy unit. The electronic density is half-filling
i.e. one electron per site. The actual value of the lattice parameter is not important in our qualitative study.

While in principle a gauge field in the hopping term should be added once magnetic fields are included, the effects we focus on occur even for very small fields. It would be confusing to the readers to alter simultaneously the geometry and the Hamiltonian of the dice model~\cite{wang11}. Thus, including gauge fields is postponed for when concrete physical realizations of dice planes or ladders are experimentally studied in materials or cold atoms.

The Rashba-SOC term in the Hamiltonian is
\begin{equation}
H_{SOC} = -\lambda\sum_{\substack{\mathbf{r},\mathbf{r}',\\ \alpha,\beta,\sigma,\sigma'}}\left(\mathfrak{i} c^{\dagger}_{\mathbf{r},\alpha,\sigma}(\hat{D}_{\alpha\beta}.\vec{\tau})_{\sigma\sigma'}c_{\mathbf{r}',\beta,\sigma'}+h.c.\right),
\end{equation}\label{Eqn: RSOC term}

\noindent where $\mathfrak{i}$ is the imaginary unit, $\vec{\tau}=\tau_{x}\hat{x} + \tau_{y}\hat{y} + \tau_{z}\hat{z}$ is the Pauli matrix vector, $\hat{D}_{\alpha\beta}$ is the unit vector in-plane and perpendicular to the bond formed by $(\mathbf{r},\alpha)$ 
and $(\mathbf{r}',\beta)$ (see illustration in Appendix A). 
The hopping
is only between nearest-neighbor sites.

The last term of the tight-binding Hamiltonian is the interaction of the spin of electrons with a magnetic field
\begin{equation}
H_{B} = -B\sum_{\mathbf{r},\alpha, \sigma, \sigma'}\tau^{z}_{\sigma\sigma'}c^{\dagger}_{\mathbf{r},\alpha,\sigma}c_{\mathbf{r},\alpha,\sigma'},
\end{equation}\label{Eqn: Magnetic term}

\noindent where $B$ is the strength of the magnetic field perpendicular to the plane of the lattice along the $z$-direction. This magnetic term breaks time-reversal. The Hamiltonian still has translational invariance. For typical oxides the hopping $t$ is approximately 0.1~eV, and thus the magnetic field used here of value $0.2t$, as in~\cite{wang11}, is large. However, the flat band split found in ~\cite{wang11} and here occurs for much smaller
magnetic fields. Thus, $B=0.2t$ is only used for clarity in the visualization of results. The physics does not change if much smaller fields are used. 


\begin{figure}[H]
\includegraphics[width=3in, height=2.5in]{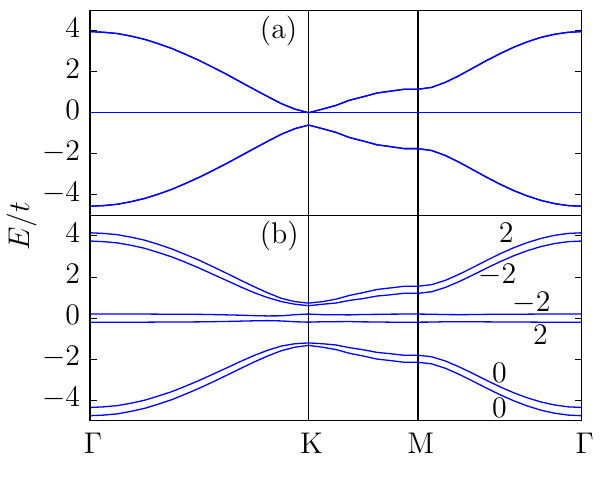}
\caption{Energy bands for the $N\times N$ dice lattice obtained diagonalizing the 6$ \times $6 Hamiltonian matrix Eq.(\ref{Eqn: Hamiltonian in k-space}) in momentum space. (a) are results for $\lambda=B=0$, while (b) are for $\lambda=0.3t$ and $B=0.2t$. In both 
cases, $\epsilon=0.6t$. These bands, already shown in~\cite{wang11}, are reproduced here for the benefit of the readers.}\label{Energy Bands NxN}
\end{figure}

In momentum space,  the non-interacting Hamiltonian becomes 
\begin{eqnarray}
&& \hspace*{2cm} H(\mathbf{k}) = \nonumber \\
&& \hspace*{-1cm}\begin{pmatrix}
-B & 0 & -t\gamma_{\mathbf{k}}^{*} & -\mathfrak{i}\lambda\gamma_{\mathbf{k}+}^{*} & 0 & 0\\
0 & B & -\mathfrak{i}\lambda\gamma_{\mathbf{k}-}^{*} & -t\gamma_{\mathbf{k}}^{*} & 0 & 0\\
-t\gamma_{\mathbf{k}} & \mathfrak{i}\lambda\gamma_{\mathbf{k}-} & -\epsilon-B & 0 & -t\gamma_{\mathbf{k}}^{*} & \mathfrak{i}\lambda\gamma_{\mathbf{k}+}^{*} \\
\mathfrak{i}\lambda\gamma_{\mathbf{k}+} & -t\gamma_{\mathbf{k}} & 0 & -\epsilon+B & \mathfrak{i}\lambda\gamma_{\mathbf{k}-}^{*} & -t\gamma_{\mathbf{k}}^{*}  \\
0 & 0 & -t\gamma_{\mathbf{k}} & -\mathfrak{i}\lambda\gamma_{\mathbf{k}-} & -B & 0 \\
0 & 0 & -\mathfrak{i}\lambda\gamma_{\mathbf{k}+} & -t\gamma_{\mathbf{k}} & 0 & B \label{Eqn: Hamiltonian in k-space}
\end{pmatrix} 
\end{eqnarray}

\noindent after the Fourier transform $c^{\dagger}_{\mathbf{k},\alpha,\sigma}=\frac{1}{\sqrt{3NM}}\sum_{\mathbf{r}}e^{\mathfrak{i}\mathbf{k}.\mathbf{r}}c^{\dagger}_{\mathbf{r},\alpha,\sigma}$ is used, with $N$ ($M$) the number of points in the long (short) direction. We defined
$\gamma_{\mathbf{k}}$=$1+e^{\mathfrak{i}k_1}+e^{\mathfrak{i}k_2}$, and $\gamma_{\mathbf{k} \pm}$=$ 1+e^{\mathfrak{i}(k_1 \pm 2\pi/3)}+e^{\mathfrak{i}(k_2 \pm 4\pi/3)}$, where the components are along the axes indicated in Fig.~\ref{fig1} 
as $k_i=\mathbf{k}.\hat{e}_{i}$. The annihilation operator basis is $(c_{\mathbf{k},1,\uparrow}, \hspace{0.15cm} c_{\mathbf{k},1,\downarrow}, \hspace{0.15cm} c_{\mathbf{k},2,\uparrow}, \hspace{0.15cm} c_{\mathbf{k},2,\downarrow}, \hspace{0.15cm} c_{\mathbf{k},3,\uparrow}, \hspace{0.15cm} c_{\mathbf{k},3,\downarrow})$. This matrix can be diagonalized and results are in Fig.~\ref{Energy Bands NxN}: (a) at $\lambda=0$ and $B=0$, there are three bands, with one totally flat and touching another at the $K$ point~\cite{wang11}; (b) with nonzero $\lambda$ and $B$ the original three bands split into six, with the Chern numbers (definition in Appendix B) as indicated.

In this first numerical investigation of electronic correlation effects on dice ladders, as described in Sec.~V, we also introduce the repulsive Hubbard interaction term which reads as
\begin{equation}
H_U = U\sum_{\mathbf{r},\alpha} n_{\mathbf{r},\alpha,\uparrow}n_{\mathbf{r},\alpha,\downarrow},
\end{equation}

\noindent
where $U$ is the strength of the Hubbard interaction,  and the number operator is $n_{\mathbf{r},\alpha,\sigma}=c^{\dagger}_{\mathbf{r},\alpha,\sigma}c_{\mathbf{r},\alpha,\sigma}$.

\subsection{Observables Calculated}


\subsubsection{Currents}
The charge current operators are defined as the sum of two terms, $J^{\text{K}}_{ij}$ and $J^{\text{SO}}_{ij}$, known as the spin preserving and spin flipping currents, respectively~\cite{Riera01}. The first has a canonical form and arises from the kinetic energy term, while the second originates in the Rashba term and vanishes if $\lambda=0$. We followed the local conservation of charge current to derive these operators. Some details of the calculation of these charge current operators are in Appendix~A and in Ref.~\cite{Riera01}.

\subsubsection{Hall Conductance}
From the currents, we computed the transverse Hall conductance ($\sigma_{xy}$), described by the Kubo formula as recently studied in Ref.~\citep{Mohanta01}: 
\begin{eqnarray}
&&\sigma_{xy}= \nonumber \\
&& \frac{2\pi}{N_s}\sum_{\varepsilon_{m}\neq \varepsilon_{n}}\frac{f_{m}-f_{n}}{\zeta^{2}+(\varepsilon_{n}-\varepsilon_{m})^{2}}\mathrm{Im}\left({J_{\hat{e}_1}}_{mn}{J_{\hat{e}_2}}_{nm}\right),\label{Eqn: Sigma_xy definition}
\end{eqnarray}

\noindent
where $J_{\hat{e}_1}$ and $J_{\hat{e}_2}$ are the current operators calculated along the lattice vectors $\hat{e}_1$ and $\hat{e}_2$, respectively. ${J_{\hat{e}_i}}_{mn}=\langle m|J_{\hat{e}_i}|n\rangle$ is the $mn^{th}$ matrix element of the current operator $J_{\hat{e}_i}=J^{K}_{\hat{e}_i}+J^{SO}_{\hat{e}_i}$. $N_s=3MN$ is the total number of sites, and $f_{m}=\left(1+e^{-\beta(\epsilon_m -\mu)}\right)^{-1}$ is the Fermi function. $\zeta$ is the relaxation parameter 
introduced to smooth the set of delta functions that arises from finite-size clusters, and to crudely represent effects not included in the model, such as lattice fluctuations. $|n\rangle$ refers to the eigenstates of the full Hamiltonian, which in most of the calculations below will be in a real-space basis.

\begin{figure}[H]
\begin{center}
\includegraphics[width=2in, height=1.2in]{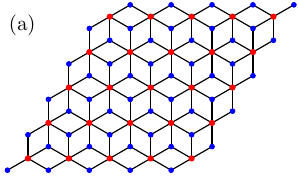}
\end{center}
\vspace*{-0.9cm}
\includegraphics[width=3in, height=2.5in]{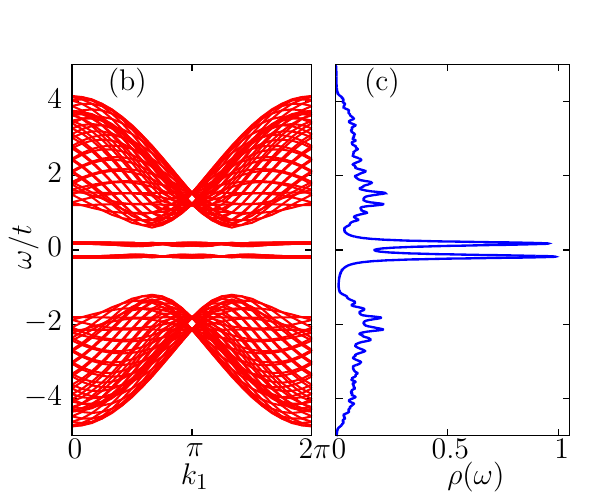}
\caption{(a) Sketch of an $N \times N$ cluster. In this illustration $N$ is 5 (focus on the red sites to count unit cells).  
(b) Bands derived from a 24$\times$24 cluster (PBC$\times$PBC) using Eq.(\ref{Eqn: Hamiltonian in k-space}), 
at $\lambda = 0.3t$ and $B=0.2t$.
Shown are 24$\times$6 bands, with the 6 arising from the product of 3 sites in the unit cell 
and a factor 2 produced by the magnetic field split. Each band has 24 points, joined by interpolating lines.
(c) DOS derived from (b) by summing over $k_1$. Note the sharp peaks near zero energy associated with near flat bands in the DOS, as it occurs in the bulk 
system in Fig.~\ref{Energy Bands NxN}~(b). The other dispersive bands are associated with the upper and 
lower bands in Figs.~\ref{Energy Bands NxN}~(a,b).
}
\label{Fig. DOS and Bands for 24x24}
\end{figure}

\subsubsection{Spectral Function}
The one-particle spectral function used in our calculation employs only momenta along the axis 1 that will 
remain ``long'' in our ladders, while axis 2 will be shorten. Specifically, we use the formula
\begin{eqnarray}
& A(k_1 &,\omega) = \frac{1}{N}\sum_{\substack{r_1,r_1',\\ r_2,\alpha,\sigma}}\sum_{m}e^{\mathfrak{i}k_1(r_1'-r_1)} \nonumber \\
 &&\Psi_{m}^*(r_1,r_2,\alpha,\sigma)\Psi_{m}(r_1',r_2,\alpha,\sigma)\delta(\omega-E_{m}),\label{arpes}
\end{eqnarray}

\noindent where $r_1$ ($r_2$), $r_1'$  are the unit cell indices along the lattice vector $\hat{e}_{1}$ ($\hat{e}_2$). Here, $k_1=\frac{2\pi n_1}{N}$ is the momentum along $\hat{e}_1$, where $n_1=0,\cdots ,N-1$, and $N$ is the number of sites along the long direction of our ladders. $m$ is the eigenvector index with energy $E_m$, and we define 
the amplitude $\Psi_{m}(r_1,r_2,\alpha,\sigma) = \langle r_1,r_2,\alpha,\sigma | m \rangle $. We provide width to the delta functions $\delta(\omega -E_n)$ via a Lorentzian function $\frac{1}{\pi}\frac{1}{(\omega-E_m)^2+\eta^2}$, with $\eta$ the broadening of this function. $\eta$ simulates effects not considered in the model, such as disorder, lattice fluctuations, and temperature, and also allows for a better visualization (smoothing) of results that otherwise would consist of sharp $\delta$ functions using a
finite cluster. Broadening was also used in our recent study of the skyrmion lattice~\cite{Mohanta01}.

\subsubsection{Spin Correlation}
To understand the ground-state properties of the Hamiltonian in the presence of interaction 
at half-filling, we calculated the real-space spin-spin correlation function between two sites 
$\langle \mathbf{S}_{i}.\mathbf{S}_{j}\rangle$ = $\langle {S}^{z}_{i}{S}^{z}_{j}\rangle + \frac{1}{2}\left(\langle {S}^{-}_{i}{S}^{+}_{j}\rangle + \langle {S}^{+}_{i}{S}^{-}_{j}\rangle\right)$.
These spin correlations were computed via both Lanczos and DMRG techniques. For Lanczos, we used the 2$\times$2 (i.e. 12-sites) cluster. 
Since the total $z$-component spin $S^z_{Total}$ is not a good quantum number due to the Rashba-SOC term in the Hamiltonian, the Hilbert-space dimension for our case at half-filling was $^{24}C_{12}=2.704156\times 10^6$ states. 
Meanwhile, the DMRG calculation was performed on the 8$\times$2 ladder system, using a maximum of 1000 states. The truncation error in the calculation of the ground state was $\sim 10^{-7}$.

\section{Results}

\subsection{$N \times N$ System}
We start with a cluster resembling the two-dimensional system but being finite
in both directions. This provides a test that working on a finite cluster can properly reproduce
the results for the bulk dice lattice~\cite{wang11}. 
We first study the eigenvalues obtained by discretizing momentum space, as it occurs for a finite system,
and later we move into real-space perspectives and calculations of relevance for this problem, such as edge currents.

In Fig.~\ref{Fig. DOS and Bands for 24x24}~(a), 
we show a small typical two-dimensional dice cluster. The dangling sites shown have indeed such a character (singular points) when OBC are used, but
for PBC in both directions all sites are equivalent i.e. blue and red dots have the same connectivity all over the lattice for PBC$\times$PBC.

In panel (b), a 24$\times$24 cluster is diagonalized. We use the perspective
of the direction ``1'' which will become the ``long'' (or leg) direction for dice ladders.
Thus, in panel (b) 24$\times$6 bands are shown, see caption. 
The DOS is in panel (c). The results are remarkably close to
those implied from Fig.~\ref{Energy Bands NxN}~(b), particularly with regards to the crucial two flat bands near zero energy. In the DOS upper and lower dispersing bands, only small spikes are the remnants of the finite-size cluster used. In
summary, a 24$\times$24 PBC$\times$PBC cluster is sufficient to capture the essence of the dice lattice, as will also be shown below with regards to several quantities.

\begin{figure}[H]
\begin{center}
\includegraphics[width=2.6in, height=1.15in]{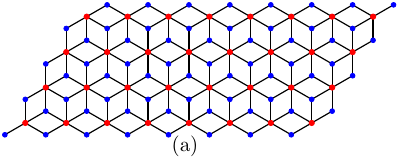}
\end{center}
\vspace*{-1.1cm}
\includegraphics[width=3in, height=2.5in]{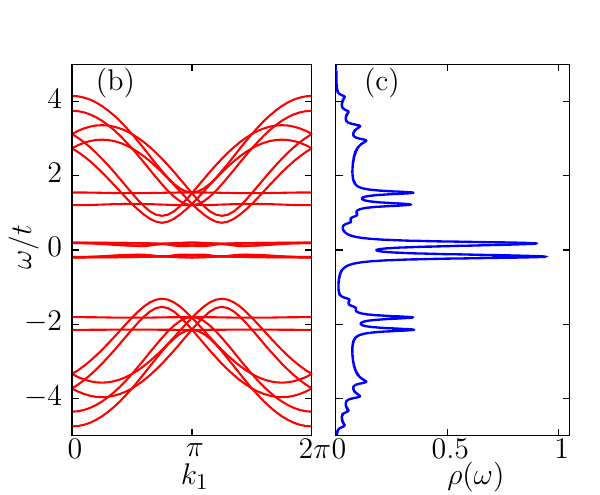}
\caption{(a) Sketch of an $N \times 4$ ladder with $N=8$ (focus on the red sites to count unit cells).  (b) Bands derived from a 150$\times$4 cluster (PBC$\times$PBC) diagonalizing Eq.(\ref{Eqn: Hamiltonian in k-space}) at $\lambda = 0.3t$ and $B=0.2t$. Shown are $4\times 6=24$ bands, with 
the factor 6 explained in the caption of Fig.~\ref{Fig. DOS and Bands for 24x24}.  (c) DOS derived from (b) by summing over $k_1$. Note the 6 sharp peaks associated with near flat bands, 
2 of which are isolated near zero energy (the important ones), 
while 4 are mixed with other non-flat bands.\label{Fig. DOS and Bands for Nx4}}
\end{figure}

\subsection{$N \times 4$ Ladder}
Consider now a dice ladder with 4 unit-cells along the short direction, as sketched 
in Fig.~\ref{Fig. DOS and Bands for Nx4}~(a). Note that 
because we have a unit cell with three sites, effectively there are 12 lines of sites, with a vertical 
periodicity blue-blue-red. This shows that numerically the effort to study these ladders after adding electronic
interactions $U$ is harder than the nominal size
suggests. Specifically, a $N \times 4$ dice ladder demands an effort similar to a  $N \times 12$ one-orbital Hubbard model ladder from the DMRG perspective.

With regards to the relevance of dice ladders to mimic planes, 
what is interesting are the results in Figs.~\ref{Fig. DOS and Bands for Nx4} (b,c). 
In spite of the very different lengths in both directions, panels (b,c) are similar to
Figs.~\ref{Fig. DOS and Bands for 24x24}~(b,c). 
In particular, both display almost identical flat bands near 
zero energy. To the extend that one of the primary non-trivial aspect of dice lattices 
are the flat bands, then the $N \times 4$ ladder keeps that aspect intact. In addition, surprisingly, there 
are other flat bands in this 4-leg ladder, 2 in the original upper-energy dispersing region and 2 in the lower-energy region.
However, these bands are not isolated from the rest, even with a nonzero $\lambda$ and $B$. Thus,
their value in inducing non-trivial properties of the system are questionable. For this reason, we will
not focus on those extra flat bands.
However, we found flat bands near zero energy, successfully confirming that a 4-leg dice ladder resembles
the two-dimensional system.

Moreover, following the procedure outlined in the Appendix based on Ref.~\cite{Fukui01} we have
calculated the Chern number of the near-zero energy flat bands of the 4-leg dice ladder. In principle,
ladders are not expected to have sharp topological properties because of their finite nature in one direction. However,
remarkably, we found $C \sim 2$ and $C \sim -2$ for the two flat bands as in Fig.~\ref{Energy Bands NxN}~(b).
For details, see below and Appendix B.
In essence, the $N \times 4$ ladder has basically the same properties as the two-dimensional dice lattice. 

\begin{figure}[H]
\begin{center}
\includegraphics[width=2.5in, height=0.95in]{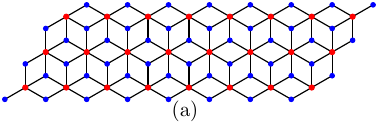}
\end{center}
\vspace*{-1.3cm}
\includegraphics[width=3in, height=2.5in]{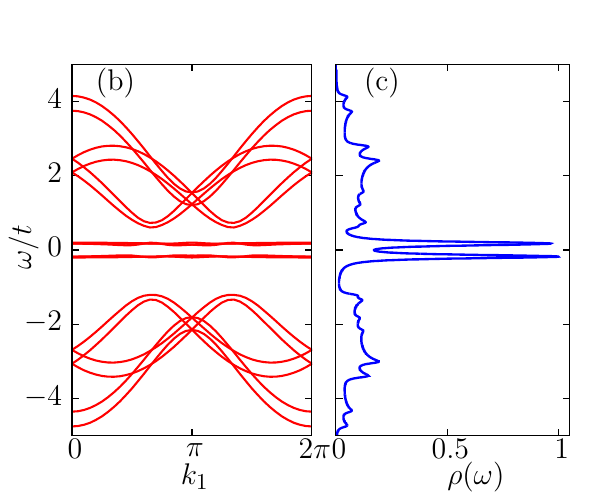}
\caption{(a) Sketch of an $N \times 3$ ladder. In the example shown $N$ is 8 (focus on the red sites to count unit cells). (b) Bands derived from a 150$\times$3 cluster (PBC$\times$PBC) diagonalizing Eq.(\ref{Eqn: Hamiltonian in k-space}) at $\lambda = 0.3t$ and $B=0.2t$. Shown are 18 bands i.e. 3$\times$6, as explained in the caption
of Fig.~\ref{Fig. DOS and Bands for 24x24}. (c) DOS derived from (b) by summing over $k_1$, clearly displaying
flat bands near zero energy.\label{Fig. DOS and Bands for Nx3}}
\end{figure}

\subsection{$N \times 3$ System}

Consider now dice ladders of size $N \times 3$, as in Fig.~\ref{Fig. DOS and Bands for Nx3}~(a). 
Here, the effective number of ``legs'' of the 3-leg dice ladder, of relevance 
for computational work in real-space when adding Hubbard $U$,
is 9 because each unit cell has 3 sites. As for the 4-leg ladder, in panel (b) we present
momentum-space results using $k_1$ in the horizontal axis (long direction). This provides 
a discrete set of bands because of the finite number of unit cells along the short direction (see caption). 
Once again, the flat bands near $E=0$ remain solidly in place. 
The DOS is still qualitatively similar to the two-dimensional
dice lattice cluster Fig.~\ref{Fig. DOS and Bands for 24x24}~(c). In this case, 
there are no extra flat bands besides those close to Fermi energy zero, contrary to 4-legs, suggesting that the existence of these extra features may depend on whether we have an even or odd number of unit cells along the short direction.

\begin{figure}[H]
\begin{center}
\includegraphics[width=2.5in, height=0.75in]{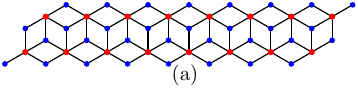}
\end{center}
\vspace*{-1.0cm}
\includegraphics[width=3in, height=2.5in]{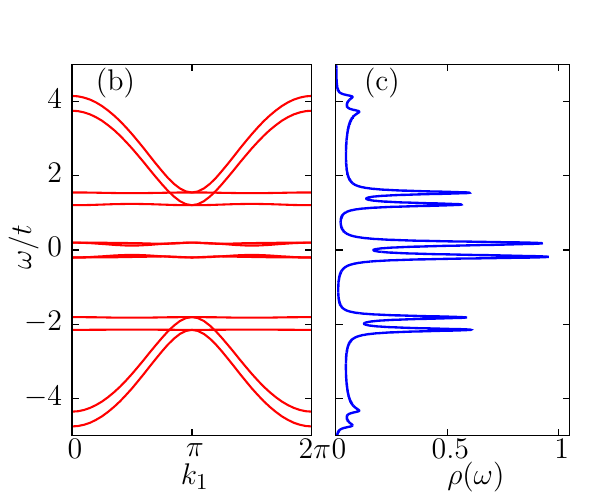}
\caption{(a) Sketch of an $N \times 2$ ladder. In the example shown $N$ is 8 (focus on the red sites to count unit cells).
(b) Bands derived from a 150$\times$2 cluster (PBC$\times$PBC)  diagonalizing
Eq.(\ref{Eqn: Hamiltonian in k-space}) at $\lambda = 0.3t$ and $B=0.2t$. Shown are 12 bands i.e. 
2$\times$6, as explained in the caption of Fig.~\ref{Fig. DOS and Bands for 24x24}. (c) DOS derived from (b) by summing over $k_1$. Note the 6 sharp peaks in the DOS associated with near flat bands, 2 of which are isolated near zero energy, while 4 are mixed with other non-flat bands.\label{Fig. DOS and Bands for Nx2}}
\end{figure}

\subsection{$N \times 2$ System}
Finally, we arrived to the ``shortest'' ribbon studied here: the 2-legs dice ladder. This system, when considered in terms
of the number of legs along the short direction, has 6 legs in the one-orbital Hubbard analog,
as illustrated in Fig.~\ref{Fig. DOS and Bands for Nx2}~(a). Thus, this system is accessible
to present computational efforts using DMRG. For this reason, the 2-leg ladder is particularly important.

In Fig.~\ref{Fig. DOS and Bands for Nx2}~(b) the bands plotted vs $k_1$ are shown, as before for 3- and 4-leg ladders. Once again, the flat bands near energy zero are clearly present, 
even in this ``shortest'' system studied here. Similarly as with 4 legs, extra flat bands were identified immersed in the upper- and lower-energy regions, but they are not of our interest. The DOS in Fig.~\ref{Fig. DOS and Bands for Nx2}~(c)
now are different from the two-dimensional cluster Fig.~\ref{Fig. DOS and Bands for 24x24}~(c) 
with regards to the upper and lower energy regions,
due to the extra flat bands. But with regards to the region near zero energy, all systems $N \times N$, $N \times 4$, 
$N \times 3$, and $N \times 2$, behave similarly: they all have similar flat bands in this energy region.

\section{Edge Currents and Edge States}
\begin{figure}[H]
\centering
\includegraphics[width=3.6in, height=3in]{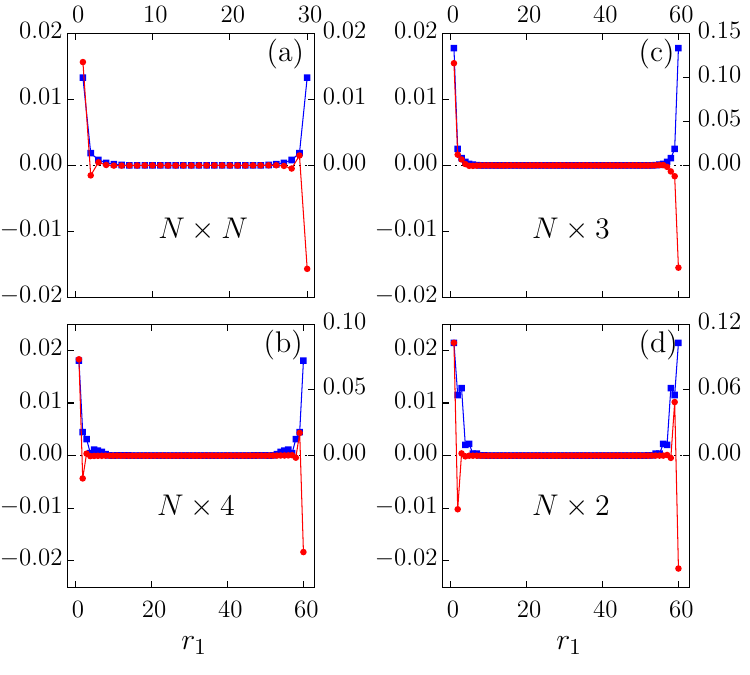}
\caption{Charge current $\langle J_{\hat{e}_2} \rangle$ (red dots, left axes) in an OBC$\times$PBC system shown vs the unit cell index $r_1$ along the long axis $\hat{e}_1$, for different dice ladders. Electronic density (blue dots, right axes) corresponding to the in-gap states i.e. states between the two flat bands
near zero, with the gap among them generated by the magnetic field. These in-gap states are induced by using cylindrical boundary
conditions. The density result is an average over the four in-gap states closest to zero energy. 
The location of both current and charge suggests these in-gap states are edge states.
Shown are averages over the short direction $r_2$ (in cylindrical geometries the results are 
translationally invariant along this PBC direction). Panels (a), (b), (c), and (d) are for ladder sizes $30\times 30$, $60\times 4$, $60\times 3$, and $60\times 2$, respectively. Parameters used in all panels are: $\lambda = 0.3t$, $B=0.2t$, $\epsilon=0.6t$. 
\label{edge currents cylinders}}
\end{figure}

\subsection{Cylindrical Boundary Conditions}

To study the properties of the $U=0$ dice ladders we performed several other 
calculations besides the DOS and their flat bands. For example, using $cylindrical$ boundary conditions, 
PBC along the short direction and OBC along the long direction, we searched for edge currents at the edges of this geometry.
Results are in Fig.~\ref{edge currents cylinders}. At the center of the cylinders -- 
away from the edges -- there are no currents as expected, but at the edges currents develop with opposite senses of circulation. There is a clear 
similarity, both qualitatively and quantitatively, between the $N \times N$ cluster that
mimics the two-dimensional plane, panel (a), and the dice ribbons in panels (b,c,d): they all behave similarly, 
with edge currents circulating in the same direction and of the same magnitude. Thus, the edge currents of two dimensional dice lattices in cylinders also appear in the  dice ladders.

\begin{figure}[H]
\hspace*{-0.5cm}
\includegraphics[width=3.33in, height=3in]{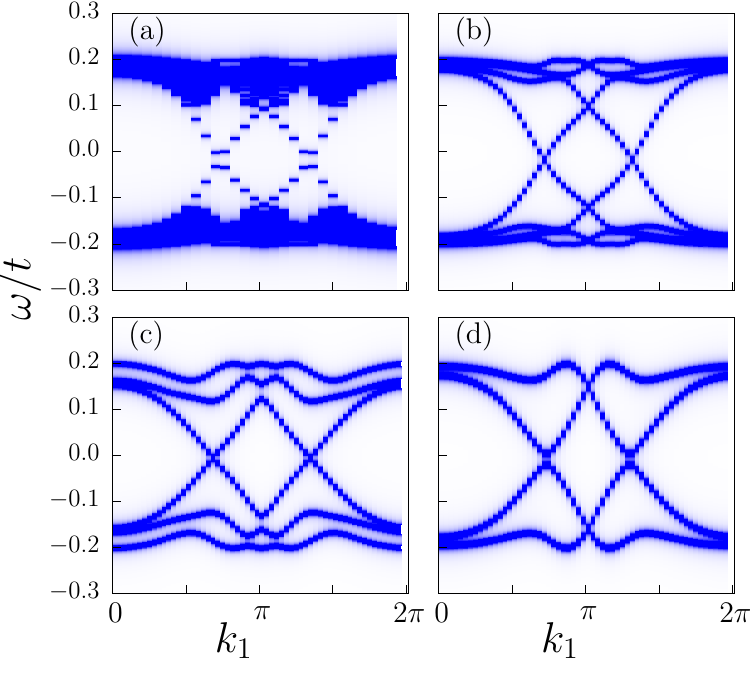}
\caption{One-particle spectral function $A(k_1,\omega)$ vs $k_1$ for various clusters, using PBC$\times$OBC (PBC in the long direction). For all panels, we used broadening $\eta=0.002t$ and energy resolution 
$d\omega=0.001t$. 
Parameters for (a) are $\lambda=0.3t$, $B=0.2t$, and $\epsilon=0.6t$, and cluster size $30\times 30$. In (b) parameters are 
$\lambda=0.3t$, $B=0.2t$, and $\epsilon=0.6t$, with cluster size $60\times 4$. In (c), parameters are $\lambda=0.7t$, $B=0.2t$, and $\epsilon=0.6t$, with cluster size $60\times 3$. In (d), parameters are $\lambda=0.5t$, $B=0.2t$, and $\epsilon=0.6t$, with cluster size $60\times 2$.\label{Fig: Akxw and HC_xy in PBCxOBC}}
\end{figure}

Using cylindrical boundary conditions creates in-gap states in between the flat bands, whose gap was opened by $B$. Studying the wave function of those in-gap states, we calculated the electronic density with 
results also in Fig.~\ref{edge currents cylinders}. 
As expected, these in-gap states are indeed
edge states, associated with the edge currents.

In Fig.~\ref{Fig: Akxw and HC_xy in PBCxOBC}, we show bands obtained from the one-particle spectral 
function $A(k_1,\omega)$ 
(technical aspects in observables Sec.~II.B) for PBC$\times$OBC, where the long direction has PBC. 
We focus on the energy range where the flat bands develop near zero energy.
For PBC$\times$PBC, these figures simply display two nearly flat bands with no extra features due to the absence of edges. 
But with PBC$\times$OBC in our finite but large clusters we observe  the 
development of edge states that manifest as crossing points. The similarity between the $N \times N$ cluster [panel (a)] 
and all the rest of the ladders [panels (b,c,d)] is clear.
In all cases, the presence of two crossing points
in principle indicates Chern number $C=2$. As explained before,
for the dice ladders we do not expect a perfect quantization to an integer of the Chern number 
calculation due to the few
momentum-space points available along the short direction. However, the conclusion that all cases panels (a,b,c,d) share
similar features is clear.
All these results suggest that two chiral edge states exist both in the two- and one-dimensional 
dice systems, even though $C$ is not quantized for ladders.

For completeness, and as with the edge states for cylinders, note that for the 3- and 2-legs dice ladders 
using other sets of $\lambda$ and $B$ parameters, we found that a small gap opens at zero energy where 
the crossings appear in Figs.~\ref{Fig: Akxw and HC_xy in PBCxOBC} (c,d) (for
4-legs these gaps are negligible). In the presence of 
those gaps, edge states were nevertheless observed for 3- and 2-legs suggesting that there is 
an effective mass in the associated chiral modes. This subtle matter will be investigated in future work.

\onecolumngrid

\begin{figure}[H]
\centering
\includegraphics[width=5in, height=2.9in]{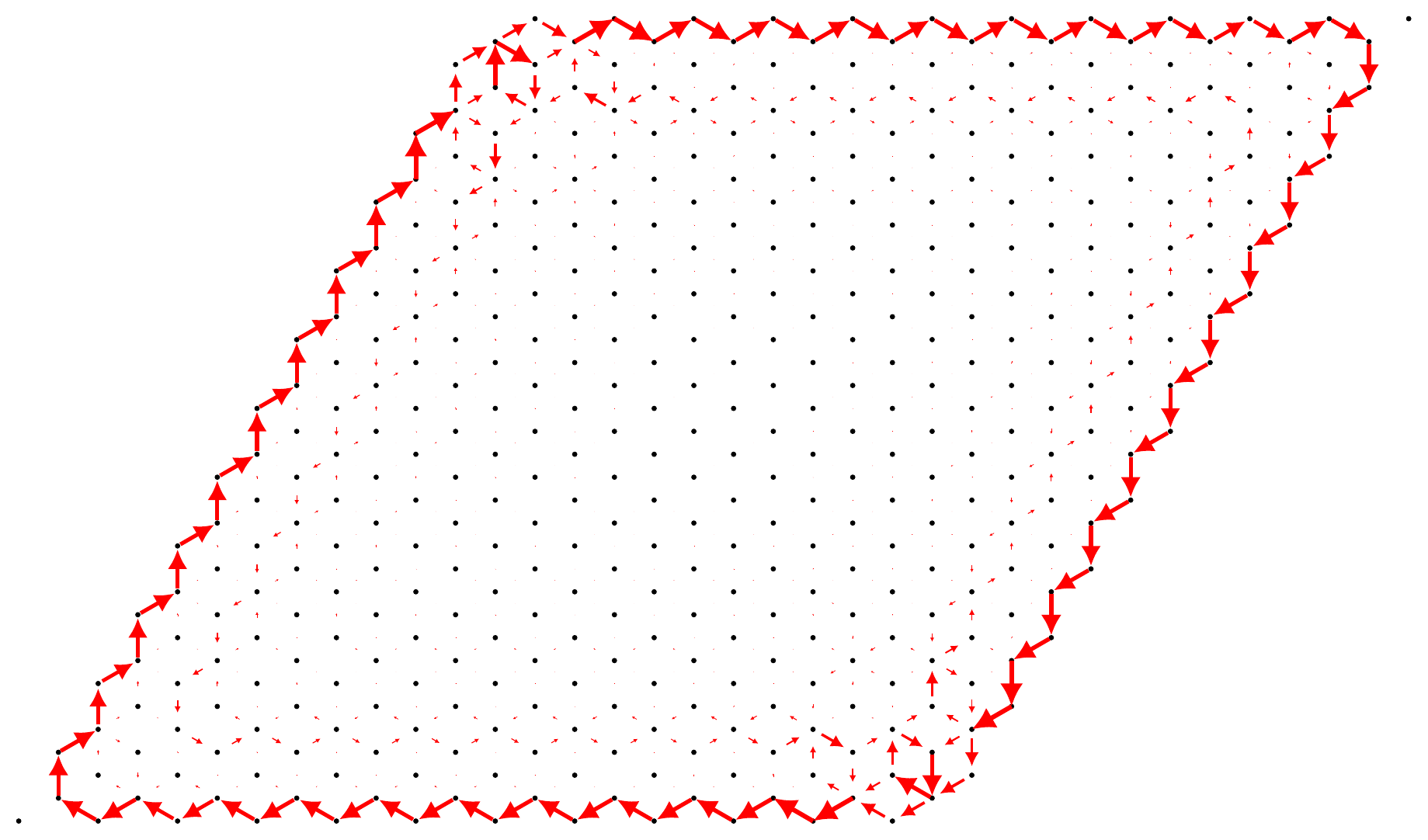}
\caption{Charge current calculated at each lattice bond on a 12$\times$12 system with OBC$\times$OBC boundary conditions, 
at $\lambda=0.3t$, $B=0.2t$, and $\epsilon=0.6t$. The presence of edge currents is obvious to the eye. The intensity of current
is proportional to the length (and also to the width) of the arrows.\label{Fig: CC 12x12}}
\end{figure}
\twocolumngrid

\onecolumngrid

\begin{figure}[H]
\centering
\includegraphics[width=6.25in, height=1.17in]{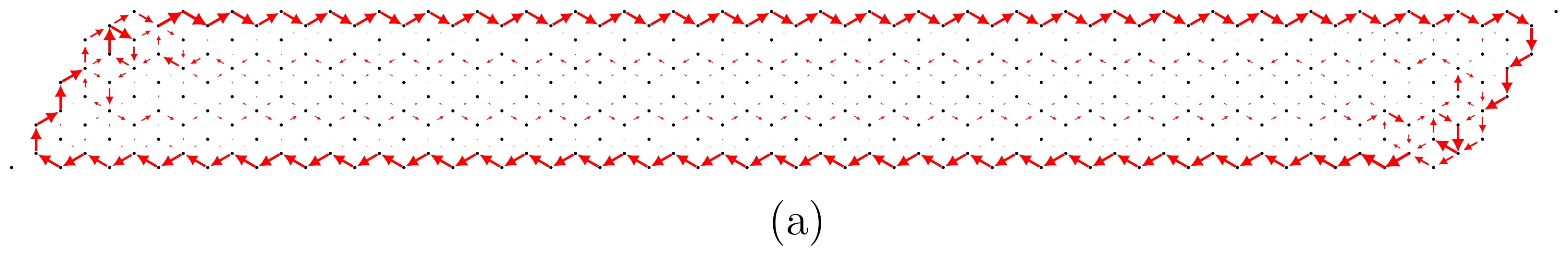}\\
\includegraphics[width=6.25in, height=0.97in]{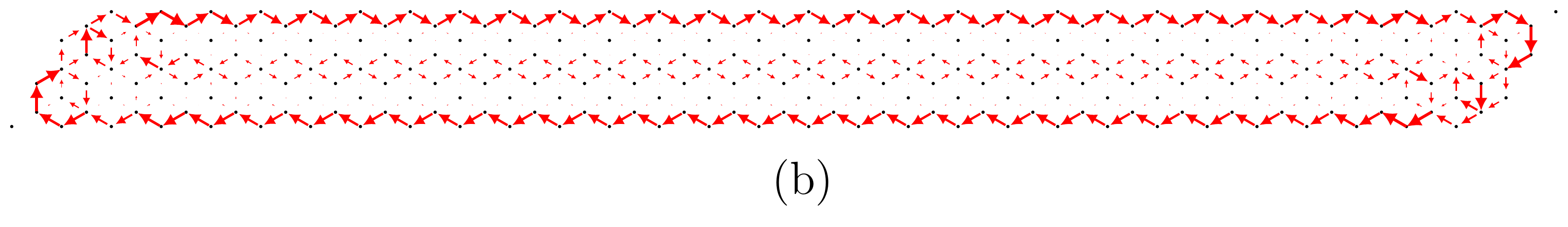}\\
\includegraphics[width=6.25in, height=0.77in]{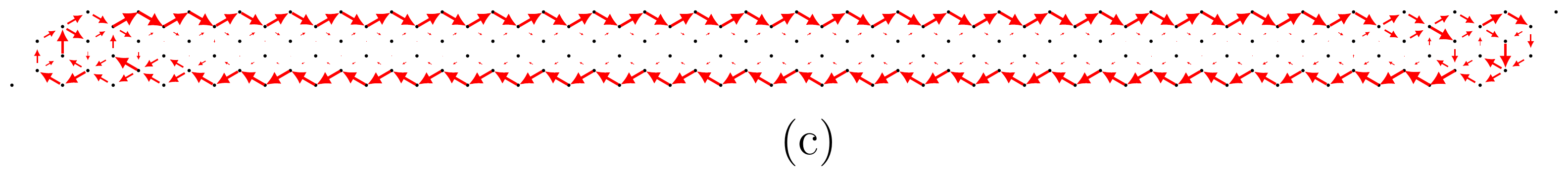}
\caption{Charge current calculated at each lattice bond for different ladder sizes. In (a) the system size is 30$\times$4 while the parameters chosen are $\lambda=0.3t$, $B=0.2t$, and $\epsilon=0.6t$. In (b) the system size is 30$\times$3, with
parameters $\lambda=0.3t$, $B=0.25t$, and $\epsilon=0.6t$. In (c) the system size is 30$\times$2 with parameters $\lambda=0.1t$, $B=0.4t$, and $\epsilon=0.6t$. All plots are using boundary conditions OBC$\times$OBC. Edge currents are present in the three ladders.\label{Fig: CC NxM}}
\end{figure}
\twocolumngrid

\subsection{Open Boundary Conditions in Both Directions}

\subsubsection{Edge Currents in $N \times N$ System}

As further test of the finite-size dice ribbons,
in Fig.~\ref{Fig: CC 12x12} we calculated the charge current for OBC along both directions, with both nonzero $\lambda$ and $B$. First, we use a cluster mimicking two dimensions, with the same number of unit cells along both directions. In Fig.~\ref{Fig: CC 12x12} edge currents develop for a 12$\times$12 cluster,
as expected. We confirmed the same for up to 30$\times$30 clusters, 
but displaying the smaller system 12$\times$12 provides better clarity.
The dangling sites of the OBC$\times$OBC geometries have zero current due to current conservation. 
In fact, we confirmed explicitly charge conservation at every site. In the two vertices without 
dangling sites, where the current changes direction abruptly, this current extends more into the bulk than at the straight edges. 

\subsubsection{Edge Currents in $N \times 4$, $N \times 3$, $N \times 2$ dice ladders}

For the 4-, 3-, and 2-leg dice ladders, results are similar as in the symmetric $12 \times 12$ cluster: in all cases 
there are sharply-defined edge currents, see Fig.~\ref{Fig: CC NxM}. As we reduce the number of legs we adjusted $\lambda$ and $B$ to better achieve the confinement of current to the edges. Without this adjustment, 
the edge currents moving left and right, along the two horizontal edges, develop an
overlap because they have a finite width near the edge. 
However, even if in some regions of parameter space the edge currents are not 
as crisp as in Fig.~\ref{Fig: CC NxM} due to overlaps,  

\begin{figure}[H]
\centering
\includegraphics[width=3in, height=2in]{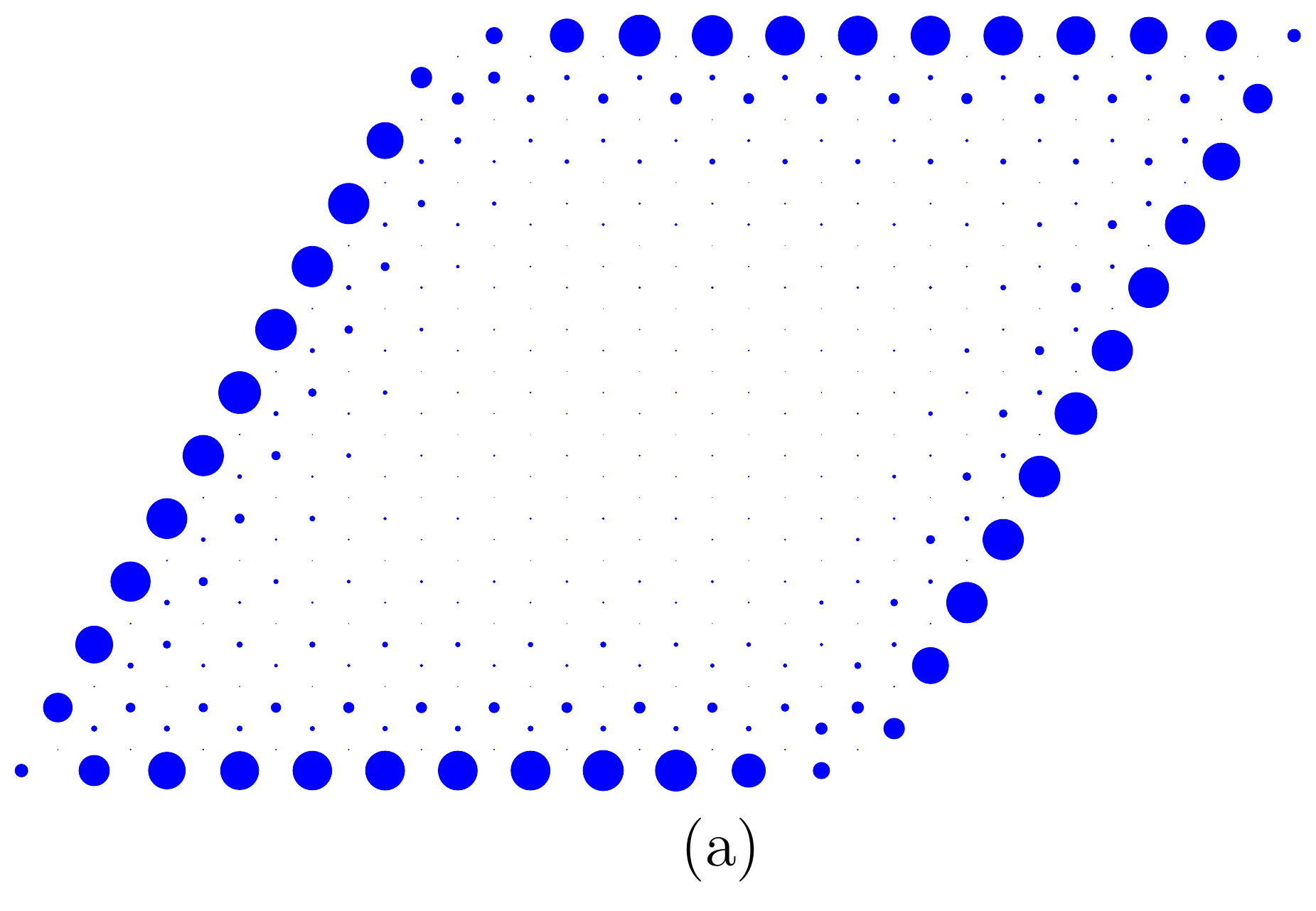}\\
\includegraphics[width=3.5in, height=0.75in]{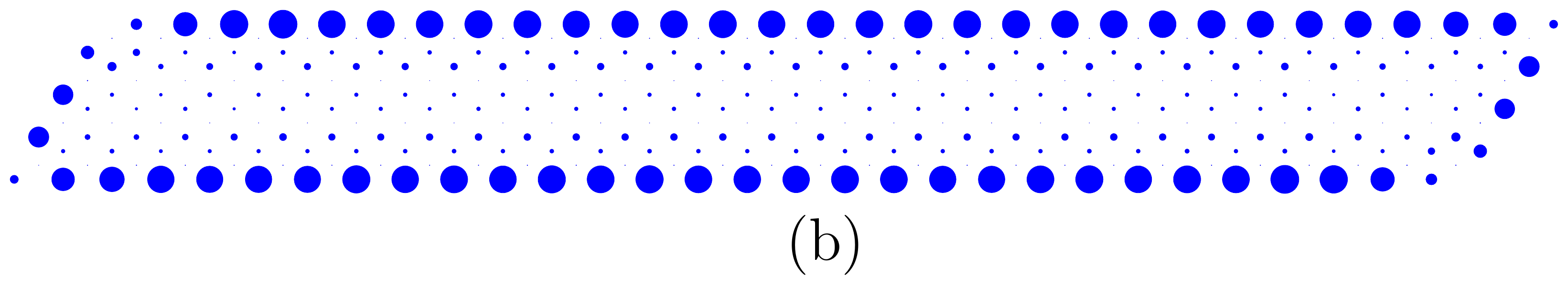}\\
\includegraphics[width=3.5in, height=0.65in]{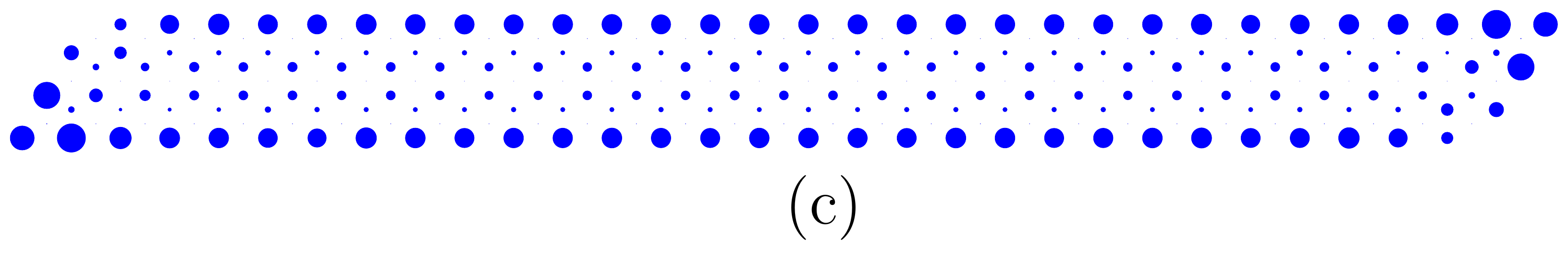}\\
\includegraphics[width=3.5in, height=0.55in]{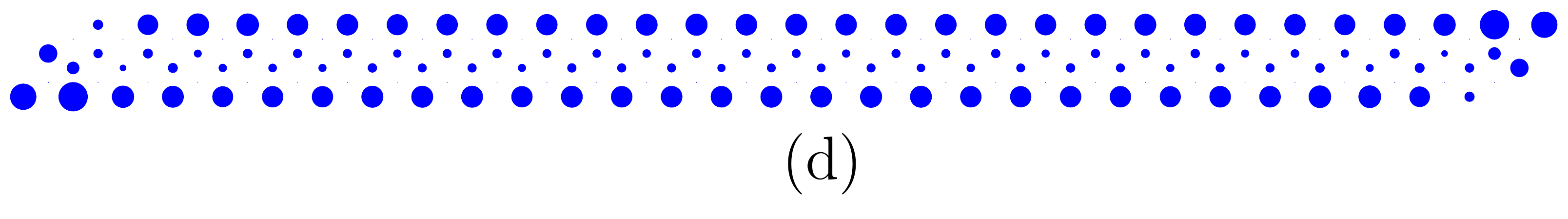}\\
\caption{Electronic density adding the contributions from all the ``edge states'' generated by using OBC$\times$OBC --in between 
the two flat bands near zero energy -- more specifically within the range 
$\Delta \omega =0.2t$ from $\omega = -0.1t$ to $0.1t$. The radius of each circle is proportional to the density 
value. 
These states, that also generate the edge currents, are localized at the edge, as expected.
Parameters used are: (a) $12 \times 12$ cluster, $\lambda=0.3t$, $B=0.2t$, and $\epsilon=0.6t$;
(b) $30 \times 4$, $\lambda=0.3t$, $B=0.2t$, and $\epsilon=0.6t$;
(c) $30 \times 3$, $\lambda=0.7t$, $B=0.2t$, and $\epsilon=0.6t$;
(d) $30 \times 2$, $\lambda=0.5t$, $B=0.2t$, and $\epsilon=0.6t$.
\label{newfig11}}
\end{figure}

\noindent
in all cases the currents sense of rotation is the same and their existence is clear.
In summary, all ladders behave similarly to planes when OBC conditions are used along both directions: in all
cases edge currents develop.

Together with the presence of currents at the edges, we also observed the presence of charge related to 
in-gap states localized at the edges. Figure~\ref{newfig11} shows the electronic density corresponding 
to the range of energies inside the gap opened by $B$ between the two flat bands.
As for the two-dimensional lattice, panel (a), 
for the dice ribbons, panels (b-d), the in-gap states generated by using OBC$\times$OBC are indeed confined to be 
``edge states'' as manifested by the edges' localization of charge.

\onecolumngrid

\begin{figure}[H]
\centering
\includegraphics[width=6.5in, height=2in]{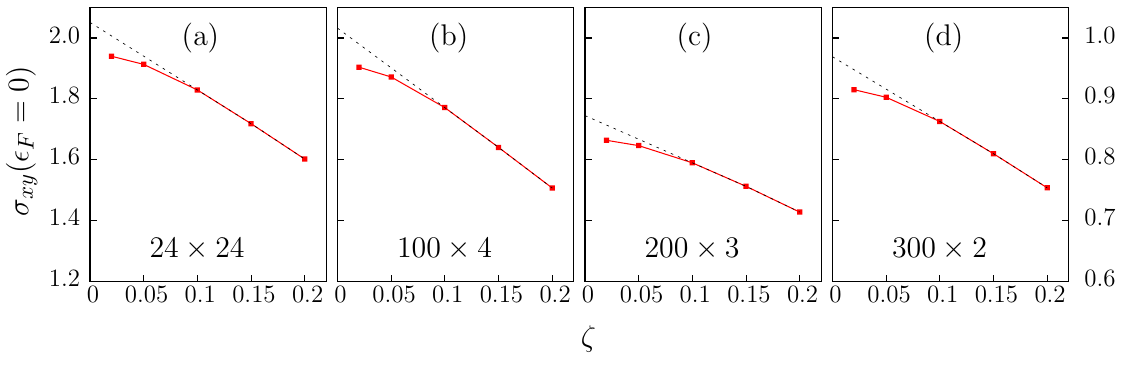}
\caption{Hall conductance ($\sigma_{xy}$) at the Fermi level ($\epsilon_{F}=0$) vs the relaxation parameter $\zeta$ for PBC$\times$PBC clusters of sizes (a) 24$\times$24, (b) 100$\times$4, (c) 200$\times$3, and (d) 300$\times$2, all for the same parameters $\lambda=0.3t$, $B=0.2t$, and $\epsilon=0.6t$. (a,b,c) share the vertical axis on the left, while (d) has its own vertical axis on the right. The extrapolation to zero broadening can be performed using all the red dots data or instead the dashed-line shown
employing only broadenings above $0.1t$. The reason is that the systematic development 
of a negative curvature in the data, deviating from
a straight line, could be a size or broadening effect. In panel (a) clearly both extrapolations via all the red dots or only
using those leading to the dashed line suggest a convergence very close
to the expected $C=2$.\label{Fig: HC_xy vs eta in PBCxPBC}}
\end{figure}
\twocolumngrid

\subsubsection{Hall conductance}

The Hall conductance was also calculated. Typically $\sigma_{xy}$ grows with increasing
energy, starting from negative energy when the first flat band is reached. This is to be expected because
of the nonzero value of $C$ for the flat bands near zero energy in two dimensions. 
For PBC$\times$PBC and after entirely crossing the first lower flat band, $\sigma_{xy}$ remains constant, and then it decreases when reaching the second band that has the same $C$ but of opposite sign.  
For OBC$\times$OBC or PBC$\times$OBC, the edge states levels in between the flat bands makes $\sigma_{xy}$  to evolve more smoothly in  value in between the flat bands, as opposed to having a plateaux. 
However, overall all the clusters studied, both the symmetric $N \times N$ and the dice ladders, share a qualitative similar behavior in the Hall conductance. 
Note that our study of the Hall conductance is not with the goal of proposing specific experiments, but theoretically to observe how close the results for one 
dimensional ladders resemble the two dimensional results, where $\sigma_{xy}$ is proportional to $C$.

With respect to the specific value reached at the central energy $\epsilon_{F}=0$, for the $24 \times 24$ cluster and PBC$\times$PBC the number is approximately 1.9, close to the expected 2. Size effects are to be expected
and extrapolating to large $N$, we conclude that $C$ for the $N \times N$ cluster is indeed very close to 2 within our
accuracy. For the dice ladders, systematically the zero energy Hall conductance is smaller, although they are all qualitatively similar. In all cases, the $\epsilon_{F}=0$ value is larger than 1.

In Fig.~\ref{Fig: HC_xy vs eta in PBCxPBC} the Fermi-energy Hall conductance is shown vs broadening for PBC$\times$PBC where size effects are less severe. In the caption, two alternative extrapolations to zero broadening are 
discussed. For the 24$\times$24, it appears that $C=2$
is a solid conclusion, in agreement with expectations. For the 4-leg ladder, results also suggest $C \sim 2$. As explained before, we do not
expect $C$ to behave exactly similarly in dice ribbons as in two-dimensional dice lattices, 
but the results shown are sufficient to illustrate that the 4-leg
ladder is virtually identical to the two-dimensional system. However, for the 3-leg dice ladders, 
deviations are pronounced and it seems unlikely $C$ will extrapolate to 2. 
Similarly, the case of the 2-leg ladder is even more pronounced, with a $C$ close to 1.
For 3- and 2-leg ladders, the 
edge currents moving in opposite directions may start overlapping, distorting the effects found
in planes. Yet, all of the dice lattices we studied have systematic behavior similar to one another.


\section{Correlation Effects}

In this section we address the second primary goal of this publication, the effect of Hubbard correlations $U$.
As explained before, considering correlation effects in two dimensions is difficult without using severe approximations. For this reason in the first portion of this publication we verified that ribbons 
of the dice lattice (``ladders'') maintain the essential properties of the 2D version, 
such as flat bands. In this section now we introduce correlation effects $U$
using the accurate Lanczos and DMRG techniques. For the latter we employ the $8 \times 2$ cluster. Note that this cluster has a total of 48 sites and in the short direction there are 6 rows of sites. In spite of this complexity, it can still be studied with DMRG. Other dice ladders like $N \times 3$ are more computationally 
challenging and its analysis 
will be attempted in future work. However, the $N \times 2$ cluster is here shown to be sufficient to illustrate the development of non-trivial magnetic order in the dice systems with increasing $U$.

\begin{figure}[H]
\hspace*{-0.5cm}
\includegraphics[width=3.33in, height=2.8in]{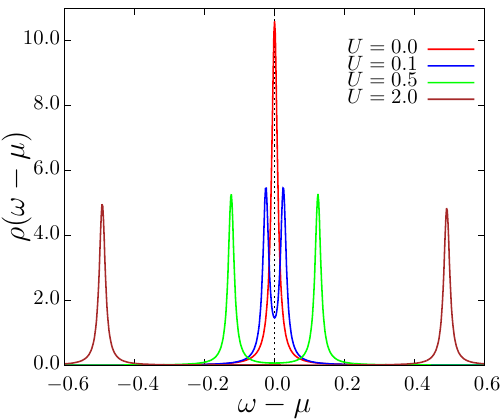}
\caption{DOS of a 2$\times$2 cluster (12 sites) studied exactly with Lanczos for various 
values of $U$, using PBC$\times$PBC. Shown are results for $\lambda=0.2t$, $\epsilon=0.6t$, $B=0$, $\eta=0.01t$ and $d\omega=0.002t$.}
\label{fig14}
\end{figure}

First, consider the effect of $U$ on the DOS at $B=0$, particularly the flat bands, using the Lanczos technique on a 2$\times$2 exactly-solved cluster (12 sites). Results are in Fig.~\ref{fig14}, focusing on the energy range near the flat bands 
of Fig.~\ref{Fig. DOS and Bands for Nx2}. From these DOS results, it is clear that $U$ splits the flat bands
similarly as a magnetic field does, with a gap growing with $U$. As explained below, the qualitative reason is that the spin state that develops as $U$ increases has ferrimagnetic tendencies, and the net global magnetization of such state qualitatively resembles an external magnetic field. 

\begin{figure}[H]
\begin{minipage}{.5\textwidth}
\includegraphics[width=2in, height=1.9in]{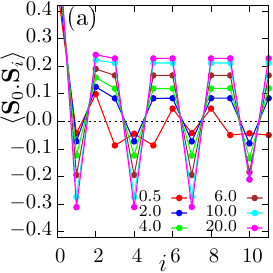}
\end{minipage}%
\begin{minipage}{.5\textwidth}
\hspace{-4cm}\includegraphics[width=1.5in, height=1.1in]{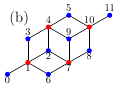}
\end{minipage}
\caption{(a) Real-space spin-spin correlation between a reference site 0 and all other sites, labeled as shown in panel (b) using the ``snake'' geometry of DMRG. We use a 2$\times$2 cluster studied with Lanczos for various $U$'s [values indicated as inset in panel (a)], PBC$\times$PBC, $\lambda=0.2t$, $\epsilon=0.6t$, and $B=0$.} 
\label{Fig: SiSj Lanczos 2x2 in PBCxPBC}
\end{figure}

In Fig.~\ref{Fig: SiSj Lanczos 2x2 in PBCxPBC}~(a), the spin-spin correlations measured exactly 
using the Lanczos technique are shown for a small cluster 2$\times$2 (12 sites), 
parametric with $U$. Site labels are in Fig.~\ref{Fig: SiSj Lanczos 2x2 in PBCxPBC}~(b). 
As the Hubbard repulsion grows, the magnetic order (the strength of 
the spin-spin correlations) grows. The pattern shown corresponds
to ferrimagnetic order, at least for $U=2$ or larger, because the number of spin correlations that are positive is double the number that are negative, with a different spin orientation for the coordination 3 and 6 sites spins. At $U=0.5$ the spin correlations are more erratic, but still there is an unbalance between positive and negative. Typically, size effects are more severe at small $U$,  but the ferrimagnetic pattern at $U=2$ or larger is robust. 

\begin{figure}[H]
\hspace*{-0.5cm}
\includegraphics[width=3.33in, height=2.5in]{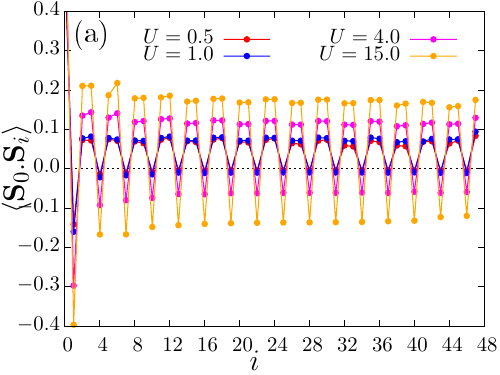}\\
\hspace*{-0.1cm}
\includegraphics[width=3.33in, height=0.9in]{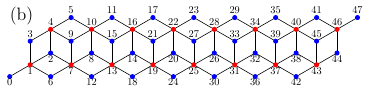}
\caption{(a) Real-space spin-spin correlation between a reference site 0 and all other sites of an 8$\times$2 cluster
(48 sites total) for different values of interaction $U$. Results shown were obtained with DMRG, and the cluster
has OBC in the long direction. Here $\lambda=0.2t$, $\epsilon=0.6t$, and $B=0$. Site labels are in panel (b).}
\label{Fig: S0Si DMRG 8x2 in OBCxOBC}
\end{figure}

The Lanczos results are in excellent agreement with the spin-spin correlations obtained using DMRG employing 
a larger 8$\times$2 cluster, as shown in Fig.~\ref{Fig: S0Si DMRG 8x2 in OBCxOBC}~(a). These results are displayed using the ``snake'' geometry where all 48 sites are arranged in a one-dimensional structure, commonly used in DMRG, see Fig.~\ref{Fig: S0Si DMRG 8x2 in OBCxOBC}~(b).
The fact that for each negative value of the correlation there are two positive is again indicative 
of the ferrimagnetic order, with a net magnetic moment, as in the Lanczos study. Note that for $U=0.5$ the ferrimagnetic
pattern is clear, suggesting that the erratic $U=0.5$ results of Lanczos are a size effect.

Thus, overall the conclusion is that accurate 
numerical techniques on dice ladders indicate that (1) ferrimagnetic order develops, and (2) the split of the flat
bands does not require an external magnetic field but correlation effects can produce related effects via ferrimagnetism. This magnetic state is in agreement with studies of the spin-1/2 Heisenberg model on a dice lattice~\cite{heisenberg}. However, our results are not consistent with those obtained with the Hartree approximation 
when including the Hubbard $U$
that predicted ferromagnetic order~\cite{wang11}, instead of ferrimagnetic, 
illustrating the value of carrying out calculations with techniques such as Lanczos and DMRG.

\begin{figure}[H]
\hspace*{-0.2cm}
\includegraphics[width=3in, height=4in]{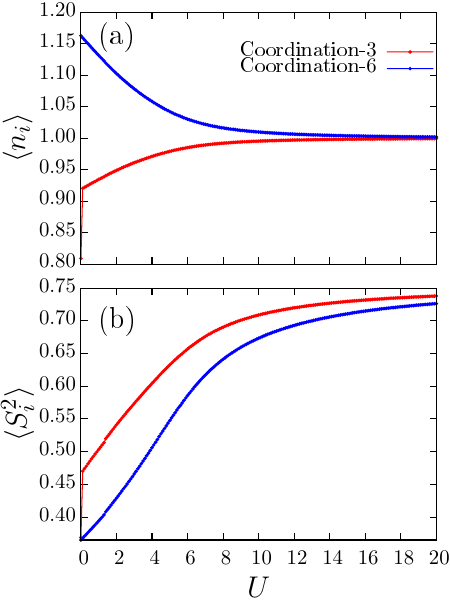}
\caption{(a) Local charge density and (b) average local moment squared for coordination-3 and coordination-6 sites in a 2$\times$2 PBC$\times$PBC system via Lanczos. The plots are for $\lambda=0.2t$, $\epsilon=0.6t$, and $B=0$.} 
\label{Fig: Local Observes Lanczos 2x2 in PBCxPBC}
\end{figure}

For completeness, we also show the Lanczos local charge density in Fig.~\ref{Fig: Local Observes Lanczos 2x2 in PBCxPBC}(a). As intuitively expected, the probability of double occupancy is strongly suppressed with increasing
$U$ and asymptotically the density 
approaches 1. Simultaneously with the local charge reaching 1, the local spin reaches 1/2
as shown in Fig.~\ref{Fig: Local Observes Lanczos 2x2 in PBCxPBC}(b) where at large $U$ the squared-spin expectation
value converges to
1/2(1+1/2)= 0.75. All this occurs similarly for coordination 3 and 6 sites, even though the weak and intermediate $U$ coupling values are slightly different.

\begin{figure}[H]
\hspace*{-0.5cm}
\includegraphics[width=3.33in, height=2.5in]{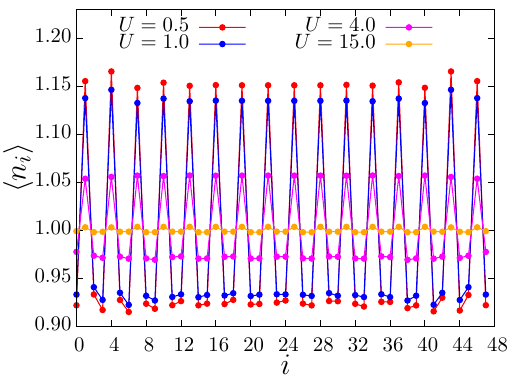}
\caption{Local electronic density for the 8$\times$2 cluster (48 sites) obtained via DMRG for different values of interaction $U$. The plots are for $\lambda=0.2t$, $\epsilon=0.6t$, and $B=0$, using the ``snake'' geometry of DMRG.} 
\label{Fig: Ni DMRG 8x2 in OBCxOBC}
\end{figure}

Similarly as with Lanczos, using DMRG we also studied the electronic density. The results in
Fig.~\ref{Fig: Ni DMRG 8x2 in OBCxOBC}, using the ``snake'' geometry, confirm that with increasing $U$ the
local densities converge to 1.

\section{Conclusions}

We have studied the dice lattice when geometrically reduced from the standard two-dimensional geometry to
quasi-one-dimensional ribbons that we call ladders. Because we aim to reach the community of experts in
correlated electronic systems, the term ``ladders'' was often used instead of ribbons by analogy to the physical systems widely studied in other contexts, such as high critical temperature superconductors~\cite{NatMatSC,PatelBaFe2S3,FeLadder7,NeutronOSMP,herbrych,ladder1,ladder2,uehara96}.
The Hamiltonian used when neglecting Hubbard repulsion is the same as in Ref.~\cite{wang11},
with a tight-binding term, Rashba spin-orbit coupling, on-site energies, and an external magnetic field. 
In addition, we incorporated the effect of Hubbard interactions
in the $N \times 2$ ladders, using Lanczos and DMRG methods.

Our main results are two folded: 

\noindent
(1) We showed that the geometrical transition from planes to ribbons 
does not alter the basic properties of the dice lattice. We arrived to this conclusion focusing 
on several non-trivial quantities of relevance in 2D, such as the flat bands
near the Fermi level, the edge currents when using OBC, the localized charge at
the edges near zero energy in the one-particle spectral function
also when using OBC, and a nonzero Hall conductance.  The main conclusion of this first portion of 
our effort is that all systems studied, more importantly the 2-leg ladder that can be studied
with DMRG including Hubbard $U$, display common properties suggesting that their qualitative behavior is  similar. 
Remarkably, the original localized states that create the flat bands are present in all the systems studied here, suggesting their localization length is very small, 
a result that could not be predicted {\it a priori}. 

\noindent
(2) In the second portion of this publication, for the first time we introduced and studied 
the effect of Hubbard correlations $U$ in dice systems using Lanczos and DMRG techniques,
for the case of the $N \times 2$ ladder.  
We show that the dice systems generate ferrimagnetic order with increasing $U$ and the flat bands split in the correlated systems without the need of introducing an external magnetic field.
Our results improve over previous mean-field efforts that reported ferromagnetic order instead.

This effort paves the way towards an in-depth study of correlation effects in dice systems. For example,
for the 2-leg dice ladder, the DOS can be analyzed in the entire $U$-$\lambda$ plane. Using DMRG such work
will require considerable computer-time resources but we estimate the effort is doable in the near future. 
We also can study how the ferrimagnetic 
order evolves increasing $\lambda$, as well as the effects of $U$ on edge currents. The future DMRG studies 
will require OBC along the long direction, as technically needed in DMRG, but the short direction can be OBC or PBC. 
All these exciting results on the challenging effects of incorporating both $U$ and $\lambda$ in dice systems 
will be presented in the near future.

\subsection*{Acknowledgments}
All authors were supported by the U.S. Department of Energy (DOE), Office of Science, Basic Energy Sciences (BES), Materials Sciences and Engineering Division. 

\section*{Appendix}

\subsection{Charge Current Operators}

\begin{figure}[H]
\centering
\includegraphics[width=3.5in, height=2in]{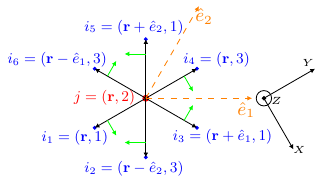}
\caption{Rashba-SOC lattice connections from the six-coordination site $j=\left(\mathbf{r},2\right)$, as example. The green arrows depict the SOC vectors $\hat{D}_{ij}$ for the bonds connected to $j$. $\hat{e}_1$ and $\hat{e}_2$ are the lattice unit vectors. $1, 2$ and $3$ are the site labels within a unit cell $\mathbf{r}$. Following~\cite{wang11}, X,Y and Z are the axes of the spin space used for the Rashba connections: the Y direction is parallel to the line from ({\bf r},2) to ({\bf r},3), and the X direction is
perpendicular.\label{Fig: Rashba-SOC connections}}
\end{figure}

From the kinetic term of the Hamiltonian and using the continuity equation for the charge,
the following current operators were derived~\cite{Riera01}:
\begin{equation}
J^{\text{K}}_{j\rightarrow i_p} = J_{j\rightarrow i_p,\uparrow \rightarrow \uparrow} + J_{j\rightarrow i_p,\downarrow \rightarrow \downarrow}
\end{equation}

\noindent
where $i_p$'s are the six neighbouring sites of site $j$ as shown in Fig.~\ref{Fig: Rashba-SOC connections} with $p=1,\cdots,6$. The current operator used above is defined as $J_{j\rightarrow i_p,\sigma \rightarrow \sigma} = -\mathfrak{i}t\left(c^{\dagger}_{i_p,\sigma}c^{\pdag}_{j,\sigma} - h.c\right)$.

Similarly, from the SOC term of the Hamiltonian, the following charge current operators were derived:
\begin{equation}
J^{\text{SO}}_{j\rightarrow i_p} = J_{j\rightarrow i_p,\uparrow \rightarrow \downarrow} + J_{j\rightarrow i_p,\downarrow \rightarrow \uparrow}
\end{equation}

\noindent
where $J_{j\rightarrow i_p, \sigma \rightarrow \sigma'} = \lambda\left(c^{\dagger}_{i_p,\sigma'}\left(\hat{D}_{i_p j}.\vec{\tau} \right)_{\sigma'\sigma}c_{j,\sigma} + h.c\right)$, and here $\sigma\neq \sigma'$.

The total charge current from site $j$ to one of the neighboring sites $i_p$ is:
\begin{equation}
J^{c}_{j\rightarrow i_p} = J^{\text{K}}_{j\rightarrow i_p} + J^{\text{SO}}_{j\rightarrow i_p}.
\end{equation}

These definitions of charge currents satisfy the conservation property (both locally and globally).

\subsection{Chern Number}
The Chern number for the $n^{th}$ band is defined in the continuum as: 
\begin{equation}
C_{n}=\frac{1}{2\pi\mathfrak{i}}\int_{T^2}F_{12}(\mathbf{k}) d\mathbf{k},
\end{equation}

\noindent
where $F_{12}(\mathbf{k})=\partial_1 A_2(\mathbf{k})-\partial_2 A_1(\mathbf{k})$ is the Berry curvature, and $A_\mu(\mathbf{k})=\langle u_{n,\mathbf{k}}|\partial_\mu|u_{n,\mathbf{k}}\rangle$ is the Berry connection. $T^2$ is the surface of a Torus formed by considering periodic boundary condition (PBC) in both the $\hat{e}_1$ and $\hat{e}_2$ directions. $\partial_\mu \equiv \partial/\partial k_{\mu}$ is the partial derivative with respect to momentum $k_1$ and $k_2$ along the $\hat{e}_1$ and $\hat{e}_2$ direction, respectively.

For lattice calculations, the above equations cannot be used unless we find Bloch functions ``$\langle \mathbf{r},\alpha,\sigma|u_{n,\mathbf{k}}\rangle$'' in a real-space basis. Since our Hamiltonian is translationally invariant, the set of Bloch states $\lbrace \langle\mathbf{r},\alpha,\sigma |\Psi_{n,\mathbf{k}}\rangle \rbrace$ are eigenstates of the Hamiltonian, where $\Psi_{n,\mathbf{k}}(\mathbf{r},\alpha,\sigma)=\langle \mathbf{r},\alpha,\sigma|\Psi_{n,\mathbf{k}}\rangle = e^{\mathfrak{i}\mathbf{k}.\mathbf{r}} u_{n,\mathbf{k}}(\mathbf{r},\alpha,\sigma)$, and in second quantization $|\mathbf{r},\alpha,\sigma\rangle = c^{\dagger}_{\mathbf{r},\alpha,\sigma}|0\rangle$. Note that the set $\lbrace\langle\mathbf{r},\alpha,\sigma|\Psi_{n,\mathbf{k}}\rangle\rbrace$ are orthogonal to each other but not necessarily the set $\lbrace \langle\mathbf{r},\alpha,\sigma|u_{n,\mathbf{k}}\rangle \rbrace$ which explains why the Berry connections are (always) non-zero.

\begin{figure}[H]
\centering
\includegraphics[width=3in, height=2.5in]{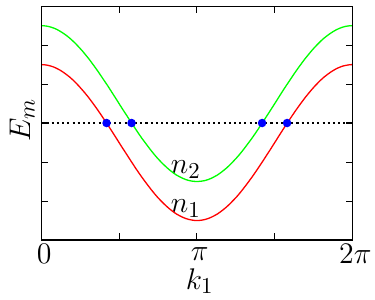}
\vspace*{-0.1cm}
\caption{$E_m$ vs $k_1$ sketch illustrating how we computed $\langle \mathbf{r},\alpha,\sigma|\Psi_{n,\mathbf{k}}\rangle$ between two non-degenerate non-touching bands (points in blue are explained in the text).\label{Fig: Additional CN}}\label{fig13}
\end{figure}

Because our computational calculation has been performed primarily in a real-space basis, the eigenstates we obtain are, in general, a superposition of all the Bloch states with the same energy, see Fig.~\ref{fig13}. For this reason, we cannot directly use the eigenstates calculated from the diagonalization computer code to find the Chern number. 

To solve this issue, consider a computer generated real-space eigenstate $|m\rangle$ with energy $E_m$. This state is a linear combination of the real-space site basis, i.e. $|m\rangle=\sum_{\mathbf{r},\alpha,\sigma} C_{\mathbf{r},\alpha,\sigma}(m)|\mathbf{r},\alpha,\sigma\rangle$, where $C_{\mathbf{r},\alpha,\sigma}(m)=\langle \mathbf{r},\alpha,\sigma|m\rangle$ are the coefficients of the transformation matrix from site basis to the real-space eigenstate of the Hamiltonian. 

As expressed above, $|m\rangle $ can also be written as a superposition of all the Bloch states $|\Psi_{n,\mathbf{k}}\rangle$ with the same energy (see Fig.~\ref{Fig: Additional CN}) as $|m\rangle = \sum_{n,\mathbf{k}} \Gamma^{n}_{\mathbf{k}}(m) |\Psi_{n,\mathbf{k}}\rangle$, where $\Gamma^{n}_{\mathbf{k}}(m)=\langle \Psi_{n,\mathbf{k}}|m\rangle$ are the transformation coefficients from Bloch states to real-space eigenstates. From the above equations and considering that $u_{n,\mathbf{k}}(\mathbf{r},\alpha,\sigma)=\langle \mathbf{r},\alpha,\sigma|u_{n,\mathbf{k}}\rangle$ are repeated between unit cells i.e. $u_{n,\mathbf{k}}(\mathbf{r},\alpha,\sigma)=u_{n,\mathbf{k}}(\mathbf{r}+\mathbf{r}',\alpha,\sigma)$, we calculate: $\sum_{\mathbf{r}'}\langle \mathbf{r}+\mathbf{r}',\alpha,\sigma|m\rangle e^{-\mathfrak{i}\mathbf{k}.\mathbf{r}'} =$
\begin{eqnarray}
&& \sum_{n,\mathbf{k}',\mathbf{r}'}\Gamma^{n}_{\mathbf{k}'}(m) \langle \mathbf{r}+\mathbf{r}',\alpha,\sigma|\Psi_{n,\mathbf{k}'}\rangle e^{-\mathfrak{i}\mathbf{k}.\mathbf{r}'} \nonumber \\
&=& \sum_{n,\mathbf{k}',\mathbf{r}'}\Gamma^{n}_{\mathbf{k}'}(m) e^{\mathfrak{i}\mathbf{k}'.(\mathbf{r}'+\mathbf{r})}\langle \mathbf{r}+\mathbf{r}',\alpha,\sigma|u_{n,\mathbf{k}'}\rangle e^{-\mathfrak{i}\mathbf{k}.\mathbf{r}'} \nonumber \\
&=& \sum_{n,\mathbf{k}',\mathbf{r}'}  \Gamma^{n}_{\mathbf{k}'}(m) e^{\mathfrak{i}\mathbf{k}'.(\mathbf{r}'+\mathbf{r})} e^{-\mathfrak{i}\mathbf{k}.\mathbf{r}'}  \langle \mathbf{r},\alpha,\sigma|u_{n,\mathbf{k}'}\rangle \nonumber \\
&=& V \sum_{n} \Gamma^{n}_{\mathbf{k}}(m) e^{\mathfrak{i}\mathbf{k}.\mathbf{r}} \langle \mathbf{r},\alpha,\sigma|u_{n,\mathbf{k}}\rangle \nonumber \\
&=& V \sum_{n} \Gamma^{n}_{\mathbf{k}}(m) \langle \mathbf{r},\alpha,\sigma|\Psi_{n,\mathbf{k}}\rangle, \label{Eqn: Bloch Functions Sum}
\end{eqnarray}

\noindent
where it must be understood that $|\mathbf{r}+\mathbf{r}',\alpha,\sigma\rangle = c^{\dagger}_{\mathbf{r}+\mathbf{r}',\alpha,\sigma}|0\rangle$, and $V$ is the volume of the system. We know that Chern numbers are well defined only for non-intersecting energy bands. In our case, with the addition of the Rashba-SOC and the Zeeman field ($B$) we fulfill this condition. From this physical argument, illustrated 
in Fig.~\ref{fig13}, we know that for fixed momentum $\mathbf{k}$ and energy $E_m$, only one value of $\Gamma^{n}_{\mathbf{k}}(m)$ is non-zero for a specific band $n$. Thus, we can write Eq.(\ref{Eqn: Bloch Functions Sum}) as:\\
\begin{equation}
 \sum_{\mathbf{r}'}\langle \mathbf{r}+\mathbf{r}',\alpha,\sigma|m\rangle e^{-\mathfrak{i}\mathbf{k}.\mathbf{r}'} =  V \Gamma^{n}_{\mathbf{k}}(m) \langle \mathbf{r},\alpha,\sigma |\Psi_{n,\mathbf{k}}\rangle.\label{Eqn: Bloch Functions Proportionality}
\end{equation}

\noindent
Of course we do not know the specific value of $\Gamma^{n}_{\mathbf{k}}(m)$, but Eq.(\ref{Eqn: Bloch Functions Proportionality}) implies that for each individual band $n$, $\langle \mathbf{r},\alpha,\sigma|\Psi_{n,\mathbf{k}}\rangle \propto \sum_{\mathbf{r'}} \langle\mathbf{r}+\mathbf{r'},\alpha,\sigma|m\rangle e^{-i\mathbf{k}.\mathbf{r'}}$, so by calculating $\sum_{\mathbf{r'}} \langle\mathbf{r}+\mathbf{r'},\alpha,\sigma|m\rangle e^{-i\mathbf{k}.\mathbf{r'}}$ and normalizing it, the Bloch states ($\Psi_{n,\mathbf{k}}(\mathbf{r},\alpha,\sigma)=\langle \mathbf{r},\alpha,\sigma |\Psi_{n,\mathbf{k}}\rangle$) can be obtained in real space for band $n$. After calculating this Bloch states, now $u_{n,\mathbf{k}}(\mathbf{r},\alpha,\sigma)$ can be obtained using $\langle \mathbf{r},\alpha,\sigma |u_{n,\mathbf{k}}\rangle = e^{-\mathfrak{i}\mathbf{k}.\mathbf{r}}\langle \mathbf{r},\alpha,\sigma |\Psi_{n,\mathbf{k}}\rangle$, which can directly be used for the calculation of the Chern number. 

Once the Bloch functions are found, we follow the method of Fukui et al.~\citep{Fukui01}, to calculate the Chern number for a discrete lattice. We first compute the U(1) link variables $U_{\mu}(\mathbf{k})=\langle u_{n,\mathbf{k}}(\mathbf{r},\alpha,\sigma)|u_{n,\mathbf{k}+\hat{\mu}}(\mathbf{r},\alpha,\sigma)\rangle/\|\langle u_{n,\mathbf{k}}(\mathbf{r},\alpha,\sigma)|u_{n,\mathbf{k}+\hat{\mu}}(\mathbf{r},\alpha,\sigma)\rangle \|$ as defined in their paper ($\hat{\mu}$ is a vector along the directions of the lattice vectors $\hat{e}_1$ and $\hat{e}_2$ with magnitude $2\pi/N_\mu$), from the Bloch functions evaluated above.
Then, we calculate the discretized lattice field strength $F_{12}(\mathbf{k})$ as:
\begin{eqnarray}
F_{12}(\mathbf{k}) &=& \partial_1 A_{2}(\mathbf{k}) - \partial_2 A_{1}(\mathbf{k}) \nonumber \\
&\approx & A_{2} (\mathbf{k}+\hat{\mu}_1) - A_{2} (\mathbf{k}) 
- A_{1} (\mathbf{k}+\hat{\mu}_2) +  A_{1} (\mathbf{k}) \nonumber \\
\end{eqnarray}

\noindent where \textit{forward} discretization of $\partial_{\mu}$ was used. Under this approximation, for example,
\begin{equation}
A_{1}(\mathbf{k})\approx \langle u_{n,\mathbf{k}}| u_{n,\mathbf{k}+\hat{\mu}_1}\rangle - \langle u_{n,\mathbf{k}}| u_{n,\mathbf{k}}\rangle .
\end{equation}

\noindent
Assuming that all $\lbrace |u_{n,\mathbf{k}}\rangle \rbrace$ are normalized to 1, it can be shown that
\begin{eqnarray}
F_{12}(\mathbf{k}) &\approx & \langle u_{n,\mathbf{k}+\hat{\mu}_1}| u_{n,\mathbf{k}+\hat{\mu}_1 +\hat{\mu}_2}\rangle - \langle u_{n,\mathbf{k}}| u_{n,\mathbf{k}+\hat{\mu}_2} \rangle  \nonumber \\
&-&  \langle u_{n,\mathbf{k}+\hat{\mu}_2}| u_{n,\mathbf{k}+\hat{\mu}_1 +\hat{\mu}_2}\rangle + \langle u_{n,\mathbf{k}}| u_{n,\mathbf{k}+\hat{\mu}_1} \rangle .
\end{eqnarray}

\begin{figure}
\includegraphics[width=3in, height=2.5in]{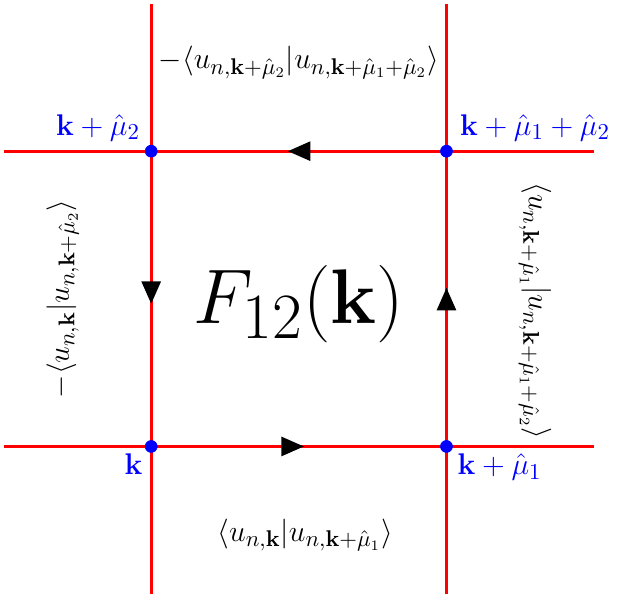}
\vspace*{-0.1cm}
\caption{Pictorial representation of the flux variable $F_{12}(\mathbf{k})$ in a $k-$space plaquette.\label{Fig: F12 Flux Plot}}
\end{figure}

Pictorially, each of these individual terms can be imagined by placing them at the links of a discretized Brillouin zone, as shown in Fig.~\ref{Fig: F12 Flux Plot}. Moreover, the sign gives a sense of circulation. $F_{12}(\mathbf{k})$ then becomes a \textit{flux} variable around an elementary plaquette of the $k-$space lattice
Then, following the directions~\citep{Fukui01} to calculate the lattice field strength $\tilde{F}_{12}(\mathbf{k})=\ln{\left( U_{1}(\mathbf{k})U_{2}(\mathbf{k}+\hat{\mu}_1)U_{1}(\mathbf{k}+\hat{\mu}_2)^{-1}U_{2}(\mathbf{k})^{-1}\right)}$, we can compute the Chern number associated with the $n^{th}$ band as:
\begin{equation}
{C}_{n}=\frac{1}{2\pi\mathfrak{i}}\sum_{\mathbf{k}}\tilde{F}_{12}(\mathbf{k}).
\end{equation}

Note that the magnitude of the quantities inside the lattice field strength $\tilde{F}_{12}(\mathbf{k})$ is always 1 and thus $-\mathfrak{i}\pi < \tilde{F}_{12}(\mathbf{k}) \leq \mathfrak{i}\pi$.

Following this momentum-space procedure, we calculated the Chern number for the $N \times N$ clusters, with focus on the flat
band below the Fermi level. The results were clearly equal to 2 (not shown). A result $C \approx 2$ 
was obtained for the $N \times 4$ ladder, confirming that these two systems are in practice virtually identical.
For the case of 3-leg ladders, however, our results seem to converge to a smaller $C$, in agreement with the deviations from
2 observed in Fig.~\ref{Fig: HC_xy vs eta in PBCxPBC}~(c). For 2-leg ladders, the calculation is not possible due to trivial
cancellations arising from the availability of only two momenta along the short direction. Nevertheless, although using 
$\sigma_{xy}$ and real-space edge currents is more reliable, the momentum-space method outlined in this Appendix provides results compatible with those in real space.

\FloatBarrier

\end{document}